\documentclass[prd,aps,twocolumn,superscriptaddress,floatfix,preprintnumbers,nofootinbib, longbibliography]{revtex4-2}

\usepackage[T1]{fontenc}
\usepackage{graphicx}   
\usepackage{dcolumn}    
\usepackage{bm}         
\usepackage{hyperref}   
\usepackage{xcolor}     
\usepackage{physics}    
\usepackage{siunitx}    
\usepackage{booktabs}   
\usepackage{microtype}  
\usepackage{subcaption}
\usepackage{ragged2e}
\hypersetup{
  colorlinks=true,
  linkcolor=blue,
  citecolor=blue,
  urlcolor=blue
}

\newcommand{\BD}[1]{{\color{orange} [BD: #1] }}

\begin{document}

\title{{\it Flash from the Past:} \\
New Gamma-Ray Constraints on Light CP-even Scalar from SN1987A}
\author{Yue Yu}
\email{yu.y1@wustl.edu}
\affiliation{Department of Physics, Washington University, St.~Louis, MO 63130, USA}
\author{Writasree Maitra}
\email{m.writasree@wustl.edu}
\affiliation{Department of Physics, Washington University, St.~Louis, MO 63130, USA}
\author{P.~S.~Bhupal Dev}
\email{bdev@wustl.edu}
\affiliation{Department of Physics, Washington University, St.~Louis, MO 63130, USA}
\affiliation{McDonnell Center for the Space Sciences, Washington University, St.~Louis, MO 63130, USA}
\author{Jean-Fran\c cois Fortin}
\email{Jean-Francois.Fortin@phy.ulaval.ca}
\affiliation{D\'epartement de Physique, de G\'enie Physique et d'Optique, Universit\'e Laval, Qu\'ebec, QC G1V 0A6, Canada}
\author{Steven P.~Harris}
\email{stharr@iu.edu}
\affiliation{Department of Physics and Astronomy, Iowa State University, Ames, IA, 50011, USA}
\affiliation{Center for the Exploration of Energy and Matter and Department of Physics,
Indiana University, Bloomington, IN 47405, USA}

\author{Kuver Sinha}
\email{kuver.sinha@ou.edu}
\affiliation{Department of Physics and Astronomy, University of Oklahoma, Norman, OK 73019, USA}

\author{Yongchao Zhang}
\email{zhangyongchao@seu.edu.cn}
\affiliation{School of Physics, Southeast University, Nanjing 211189, China} 
\affiliation{Center for High Energy Physics, Peking University, Beijing 100871, China}

\begin{abstract}
We derive new constraints on light CP-even scalars using old gamma-ray observations in the direction of SN1987A by the Solar Maximum Mission (SMM) satellite. Light scalars can be abundantly produced in the supernova core via the nucleon bremsstrahlung process, can stream out of the supernova-environment and decay into photons -- either primary photons or secondary photons from  lepton-antilepton pairs --  thus leading to a gamma-ray signal. From the non-observation of excess photon flux by SMM after the detection of the neutrino burst from SN1987A, we set new constraints on the mixing angle of the CP-even scalar with the Standard Model Higgs boson.
\end{abstract}


\maketitle

\section{Introduction}
Although the Standard Model (SM) has been quite successful in explaining various mysteries of nature at the most fundamental level, there exist compelling empirical evidence and  theoretical reasons for some beyond-the-SM (BSM) physics. Since the nature of BSM physics is largely unknown, the search for BSM particles must continue across a wide range of energies. The absence of a statistically-significant BSM signal at the Large Hadron Collider (LHC) has reignited the community interest in light, sub-GeV, weakly-interacting BSM particles~\cite{Antel:2023hkf,Arza:2026rsl}. A rich experimental program is currently underway to search for these light BSM particles at the intensity frontier~\cite{Batell:2022dpx}.

Compact astrophysical objects like stars and  supernovae provide an alternative laboratory for studying light BSM particles~\cite{Raffelt:1996wa,Fortin:2021cog}. If kinematically allowed, BSM particles can be abundantly produced in the hot, dense environment of stellar cores via their interactions with nuclei, electrons, muons, or photons. After being produced, for a certain range of their couplings to the SM, 
these BSM particles may escape the stellar core and take away appreciable energy, thus affecting stellar evolution or contradicting existing astrophysical observations. In particular, the observation of neutrinos from the type-II core-collapse supernova (CCSN) SN1987A in the Large Magellanic Cloud ushered in a new era in neutrino astronomy~\cite{Kamiokande-II:1987idp,Bionta:1987qt,Alekseev:1987ej}. The observed neutrino luminosity from   SN1987A was found to be in good agreement with theory~\cite{Bethe:1985sox, Burrows:1987zz,  Janka:2012wk} and modern simulations~\cite{Janka:2012wk, Fiorillo:2023frv, Muller:2024slv} based on the neutrino cooling mechanism~\cite{Raffelt:2025wty}. This puts stringent constraints on any additional cooling by the emission of light BSM particles. In fact, the energy transported out of the supernova core by any weakly-interacting BSM particle must not exceed a fraction of the energy transported by neutrinos -- widely known as the {\it Raffelt criterion}~\cite{Raffelt:1996wa}. The supernova cooling bound has been derived for various light BSM particles, such as CP-odd axions or axion-like particles (ALPs)~\cite{Raffelt:1987yt, Lella:2023bfb,Caputo:2024oqc}, CP-even scalars~\cite{Krnjaic:2015mbs,Hardy:2016kme,Dev:2020eam, Dev:2020jkh, 
Balaji:2022noj,Bottaro:2023gep, Yamamoto:2023zlu,
Hardy:2024gwy}, dark photons~\cite{Bjorken:2009mm,Dent:2012mx,Rrapaj:2015wgs,Chang:2016ntp,Chang:2018rso, Caputo:2025aac, Caputo:2025avc}, light $Z'$~\cite{Knapen:2017xzo,Croon:2020lrf,Manzari:2023gkt} and sterile neutrinos~\cite{Carenza:2023old}.

If the light BSM particle escapes the supernova environment and then decays to visible SM final states, we can also use the electromagnetic observations of SN1987A to derive  complementary constraints. This has been done for ALPs~\cite{Kolb:1988pe,Giannotti:2010ty,Jaeckel:2017tud, Hoof:2022xbe,Diamond:2023scc,Muller:2023vjm, Chauhan:2025xqr}, dark photons~\cite{Kazanas:2014mca,DeRocco:2019njg,Sung:2019xie,Calore:2021lih, Shin:2022ulh,Balaji:2025alr} and sterile neutrinos~\cite{Carenza:2023old, Calore:2021lih, DelaTorreLuque:2024zsr,Balaji:2025alr, Chauhan:2025mnn} using either prompt or secondary gamma-ray emission observations. However, this has not been done so far for CP-even scalars, which we aim to do here for supernova.\footnote{A similar ecosystem of studies has emerged for neutron star mergers \cite{Hook:2017psm, Huang:2018pbu, Dietrich:2019shr, Harris:2020qim, Zhang:2021mks, Dev:2021kje, Diamond:2021ekg, Fiorillo:2022piv}.}

The Gamma-Ray Spectrometer (GRS) on NASA's Solar Maximum Mission (SMM) satellite~\cite{1980SoPh...65...15F}  was operative at the time of the SN1987A explosion. Although it was pointing in the direction of the Sun, it had the ability to observe gamma-rays coming from SN1987A  through the shielding of the instrument. SMM did not register any excess gamma-ray photons over the expected background up to 223 seconds after the detection of the first neutrinos from SN1987A and thus set an upper bound on the gamma-ray fluence from the SN1987A explosion~\cite{Chupp:1989kx}. Using this information, we derive new gamma-ray constraints on light CP-even scalars decaying to photons and rule out new parameter space otherwise allowed by the existing astrophysical~\cite{Krnjaic:2015mbs,Hardy:2016kme,Dev:2020eam, Dev:2020jkh, 
Balaji:2022noj,Bottaro:2023gep, Yamamoto:2023zlu,
Hardy:2024gwy, Fiorillo:2025zzx}, cosmological~\cite{Berger:2016vxi, Fradette:2018hhl, Ibe:2021fed} and laboratory~\cite{Chen:2014oda, Dev:2017dui, Winkler:2018qyg, Batell:2019nwo, Egana-Ugrinovic:2019wzj, Dev:2019hho, Berryman:2019dme,
Batell:2022dpx, Feng:2022inv, MicroBooNE:2025gpp} constraints.  

A light CP-even scalar uncharged under the SM gauge group can arise in a variety of BSM scenarios motivated by a wide range of fundamental open questions. Depending on its mass and couplings to the SM sector, it can help stabilize the SM vacuum~\cite{Gonderinger:2009jp, Lebedev:2012zw,Elias-Miro:2012eoi,Ghorbani:2021rgs}; can play a role in addressing the neutrino mass~\cite{Dev:2017dui,Bhattacharya:2016qsg,Kim:2025zrs}, electroweak baryogenesis~\cite{Espinosa:1993bs,Choi:1993cv,Espinosa:2011ax,Beniwal:2017eik, Ellis:2022lft}, hierarchy problem~\cite{Englert:2013gz,Graham:2015cka,Flacke:2016szy}, and cosmological constant problem~\cite{Davoudiasl:2004be,Bertolami:2007wb,Foot:2011et,Dimopoulos:2018eam}; can act as a mediator of interactions between dark matter and the SM sector~\cite{Patt:2006fw,Pospelov:2007mp, Krnjaic:2015mbs, Knapen:2017xzo, 
Beniwal:2015sdl}; or be a dark matter itself~\cite{Silveira:1985rk,McDonald:1993ex,Burgess:2000yq,Casas:2017jjg,GAMBIT:2018eea}. The only renormalizable couplings of a SM-singlet CP-even scalar to the SM is via the Higgs portal~\cite{Krnjaic:2015mbs}. The corresponding interaction terms in the Lagrangian are proportional to $H^\dag H$, where $H$ is the SM Higgs field. After electroweak symmetry breaking, the singlet mixes with the Higgs, which we parametrize with the mixing angle $\sin\theta$ and expand in the mass basis to write the interactions between the singlet scalar ($S$) and the SM fermions ($f$) of the form~\cite{Krnjaic:2015mbs} 
\begin{align}
    -{\cal L} \supset \sin\theta \,  S \sum_f \frac{m_f}{v_{\rm EW}}\bar{f}f \, ,   
    \label{eq:Lag}
\end{align}
where $m_f$ is the fermion mass and $v_{\rm EW}\simeq 246.2$ GeV is the electroweak vacuum expectation value. In the simplest setup, we can  treat $\sin\theta$ and the scalar mass $m_S$ as the only free parameters in the theory, and therefore, derive our constraints in the $(m_S,\sin\theta)$ plane.  

We should keep in mind two qualitative differences between CP-even scalars and CP-odd axions (and ALPs in general), while deriving supernova bounds: \\
(i) In the nucleon bremsstrahlung process, a CP-even scalar can be emitted from any of the external nucleon legs, as well as from the pion mediator~\cite{Dev:2020eam, Dev:2021kje}; see Fig.~\ref{fig:FeynDiagram}. On the other hand, the ALPs can only be emitted from the nucleon legs, as the pseudoscalar nature of ALPs forbids the ALP-pion coupling at leading order.\\
(ii) While deriving gamma-ray constraints, ALP decay to photons ($a\to \gamma\gamma$), governed by the ALP-photon coupling $g_{a\gamma}$, is assumed to have 100\% branching ratio (BR), and  ALP decays to other possible SM final states are ignored, since they are governed by different couplings, which may (or may not) be related to the ALP-photon coupling in a model-dependent way. On the other hand, for the Lagrangian~\eqref{eq:Lag}, the scalar decay to all SM final states is governed by the same mixing angle $\sin\theta$ and, above the $e^+e^-$ threshold, the loop-induced primary photon channel $S\to \gamma\gamma$ is highly suppressed (see Appendix~\ref{App:Decay}). This requires us to consider the secondary photons from $S\to e^+e^-$, and above the muon (pion) threshold, from $\mu^+\mu^-$ ($\pi^+\pi^-$, $\pi^0\pi^0$) decay channels, which makes the photon flux calculation more involved for the CP-even scalar case than for the ALP case.

The rest of the paper is organized as follows: In Sec.~\ref{Sec:SNprod}, we describe our calculation of the scalar production rate in supernovae via the nucleon bremsstrahlung process. In Sec.~\ref{Sec:decay}, we discuss the decay and survival probability of the scalar.  Sec.~\ref{Sec:reabs} considers the reabsorption probability of the scalar in the supernova via inverse bremsstrahlung.  Sec.~\ref{Sec:SNprofile} discusses the supernova profile we have used in this work.   Sec.~\ref{Sec:results} presents the new constraint we have obtained using the SMM data. Finally, in Sec.~\ref{Sec:conclusion}, we summarize our results. In Appendix~\ref{App:Decay}, we list the relevant partial decay widths and BRs of the scalar. In Appendix~\ref{App:geometry}, we explain the decay geometry used to calculate the scalar mean free path (MFP). \BD{Rearrange at the end.}
\section{Scalar Emission from Supernova}
\label{Sec:SNprod}
 Through its mixing with the SM Higgs boson [cf.~Eq.~\eqref{eq:Lag}], the scalar $S$ couples to SM quarks, leptons and photons. Similar to the ALP case, the scalar $S$ can be produced in the supernova core via the following processes~\cite{Pantziris:1986dc}: \\
 (i) Nucleon bremsstrahlung: $NN \rightarrow NNS$,  \\
 (ii) $e^{+}e^{-}$ annihilation: $e^{+}e^{-} \rightarrow \gamma S$,\\ (iii) $\gamma\gamma$ annihilation: $\gamma\gamma \rightarrow S$, \\ (iv) Compton-like scattering: $e\gamma \rightarrow eS$, \\
 (v) Primakoff-like process: $\gamma e~(N) \rightarrow Se~(N)$,\\
 (vi) Plasmon decay: $\gamma_\text{pl} \rightarrow \gamma S$. \\
 Among these, (i) nucleon bremsstrahlung is the most efficient scalar production channel in the supernova core mainly because of the high nucleon number density, small Yukawa coupling of electrons with scalars and loop-suppressed coupling of photons with the scalar. 
 
\begin{figure*}[t!]
  \centering
  \includegraphics[width=0.4\textwidth]{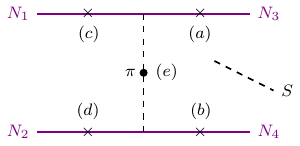}
  \hspace{0.05\textwidth}
  \includegraphics[width=0.4\textwidth]{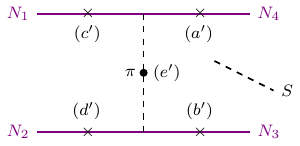}
  \captionsetup{justification=Justified}
  \caption{ Feynman diagrams for the production of light scalar $S$ in the nucleon bremsstrahlung process $N_1 + N_2 \to N_3 + N_4 +S$ in  the supernova core~\cite{Dev:2020eam}, with $N_i$ corresponding to either proton or neutron. The left and right panels are respectively showcasing the $t$ and $u$-channel Feynman diagrams of the concerned process. The light scalar $S$ can be attached to any of the nucleon lines $(a)$, $(b)$, $(c)$, $(d)$, $(a')$, $(b')$, $(c')$, $(d')$, as denoted by the crosses ($\times$), or to the pion mediator $(e)$, $(e')$, as denoted by the blobs ($\bullet$).}
  \label{fig:FeynDiagram}
\end{figure*}

An exact calculation of the scalar production rate via nucleon bremsstrahlung process in the supernova core is difficult, because the typical center-of-mass energy scale is intermediate between the validity regime of chiral effective field theory (EFT) and perturbative QCD, and also because of the interacting nature of the medium. We use the one-pion exchange (OPE) scheme, following Refs.~\cite{Dev:2020eam, Dev:2021kje}.\footnote{The validity of the OPE is discussed in Ref.~\cite{Turner:1988bt}.} The corresponding Feynman diagrams are shown in Fig.~\ref{fig:FeynDiagram}. The scalar can be emitted from one of the external nucleon legs, as shown by  $(a)$--$(d)$ for the $t$-channel and  $(a')$--$(d')$ for the $u$-channel in Fig.~\ref{fig:FeynDiagram}), with $\times$ denoting the coupling of $S$ with nucleons~\cite{Shifman:1978zn,Cheng:1988im}. In addition, the scalar can also be emitted from the (virtual) pion mediator, as shown by $(e)$ and  $(e')$ in Fig.~\ref{fig:FeynDiagram}, with $\bullet$ indicating the coupling of $S$ with pions~\cite{Voloshin:1985tc,Donoghue:1990xh}.  

The effective interaction Lagrangian for the scalar coupling with nucleons and pions is given by
\begin{equation}
-\mathcal{L}
\supset \sin\theta\, S\Big[
y_{hNN}\,\bar{N}N
+ A_{\pi}\left(\pi^{0}\pi^{0}+\pi^{+}\pi^{-}\right)
\Big] ,
\end{equation}
where $y_{hNN}\simeq 10^{-3}$ is the effective coupling of SM Higgs boson to nucleons~\cite{Shifman:1978zn,Cheng:1988im}, and $A_{\pi}$ is the effective coupling with the pion that takes the following form in chiral EFT~\cite{Voloshin:1985tc, Donoghue:1990xh}:
\begin{equation}
    A_{\pi}=\frac{2}{9v_{\rm EW}}\left(m_{S}^{2}+\frac{11}{2}m_{\pi}^{2}\right) \, .
\end{equation}
For production inside supernova, the relevant range of the scalar mass is $m_S\lesssim 2m_\pi$, where $A_\pi\simeq m_\pi^2/v_{\rm EW}$.

Following the calculations in Refs.~\cite{Giannotti:2005tn,Dev:2020eam, Balaji:2022noj}, the rate of energy emitted by scalars per unit volume in the supernova core via the nucleon bremsstrahlung process can be written as
\begin{align}
\label{Eq:rate}
Q \ = \ & \int {\rm d} \Pi_5 {\cal S} \sum_{\rm spins} |{\cal M}|^2 (2\pi)^4
\delta^4 (p_1 + p_2 - p_3 - p_4 - k_S) \nonumber \\
& \qquad \times \, \omega_S f_1 f_2 P_{\rm decay} P_{\rm escape} \,,
\end{align}
where $p_i$ is the four-momentum of the nucleon $N_i$, $k_S$ and $\omega_S$ are the four-momentum and energy of the scalar $S$, ${\rm d} \Pi_5 $ is the $2\rightarrow3$ phase space factor, ${\cal S}$ is the symmetry factor for identical particles (1 for $np$ and $1/4$ for $pp$ and $nn$),  $\mathcal{M}$ is the scattering amplitude for the nucleon bremsstrahlung process, and $f_{1,\,2}$ are the non-relativistic Maxwell-Boltzmann distributions of the two incoming nucleons in the non-degenerate limit. In supernova conditions, protons are clearly nondegenerate, but neutrons have mild degeneracy, which we neglect here for convenience \cite{Hannestad:1997gc,Fore:2019wib}.  Ref.~\cite{Hardy:2024gwy} found that accounting for the slight neutron degeneracy has little effect on the scalar production rate, given by
\begin{eqnarray}
f_{1,2}({\bf p}) \ = \ \frac{n_B}{2}\left(\frac{2\pi}{m_N T}\right)^{3/2}
{\rm{e}}^{-{\bf p}^2/2 m_N T} \,,
\end{eqnarray}
with $T$ being the temperature of the supernova core, $n_B$ the baryon number density in the supernova and $m_N$ the mass of the nucleon. Both $n_B\equiv\rho_B/m_N$ (with $\rho_B$ being the baryon density in the supernova) and $T$ are functions of the radial and temporal coordinates $(r,t)$.  
Details of the supernova profile used in this work are described in the Sec.~\ref{Sec:SNprofile}. $P_\text{decay}$ in Eq.~\eqref{Eq:rate} is the probability of the scalar to decay outside the supernova core radius $R_c$ and is discussed in Sec.~\ref{Sec:decay}. $P_\text{escape}=1-P_{\rm abs}$, where $P_{\rm abs}$ is the probability of the scalar being reabsorbed via inverse nucleon bremsstrahlung process; see  Sec.~\ref{Sec:reabs}. 

\subsection{Supernova Profile}
\label{Sec:SNprofile}
\begin{figure*}[t!]
  \centering
  \includegraphics[width=0.45\textwidth]{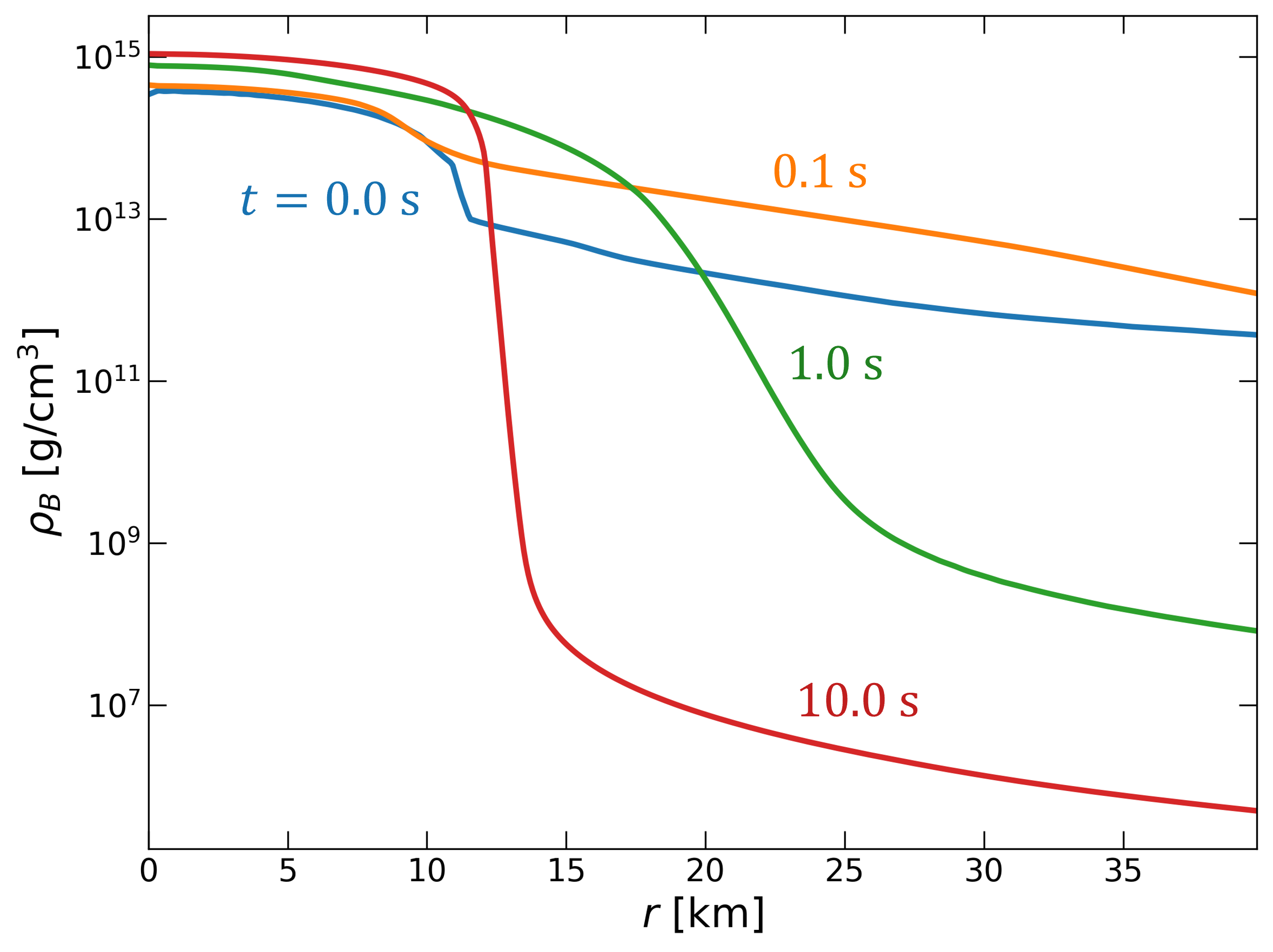}
  \hspace{0.04\textwidth}
  \includegraphics[width=0.45\textwidth]{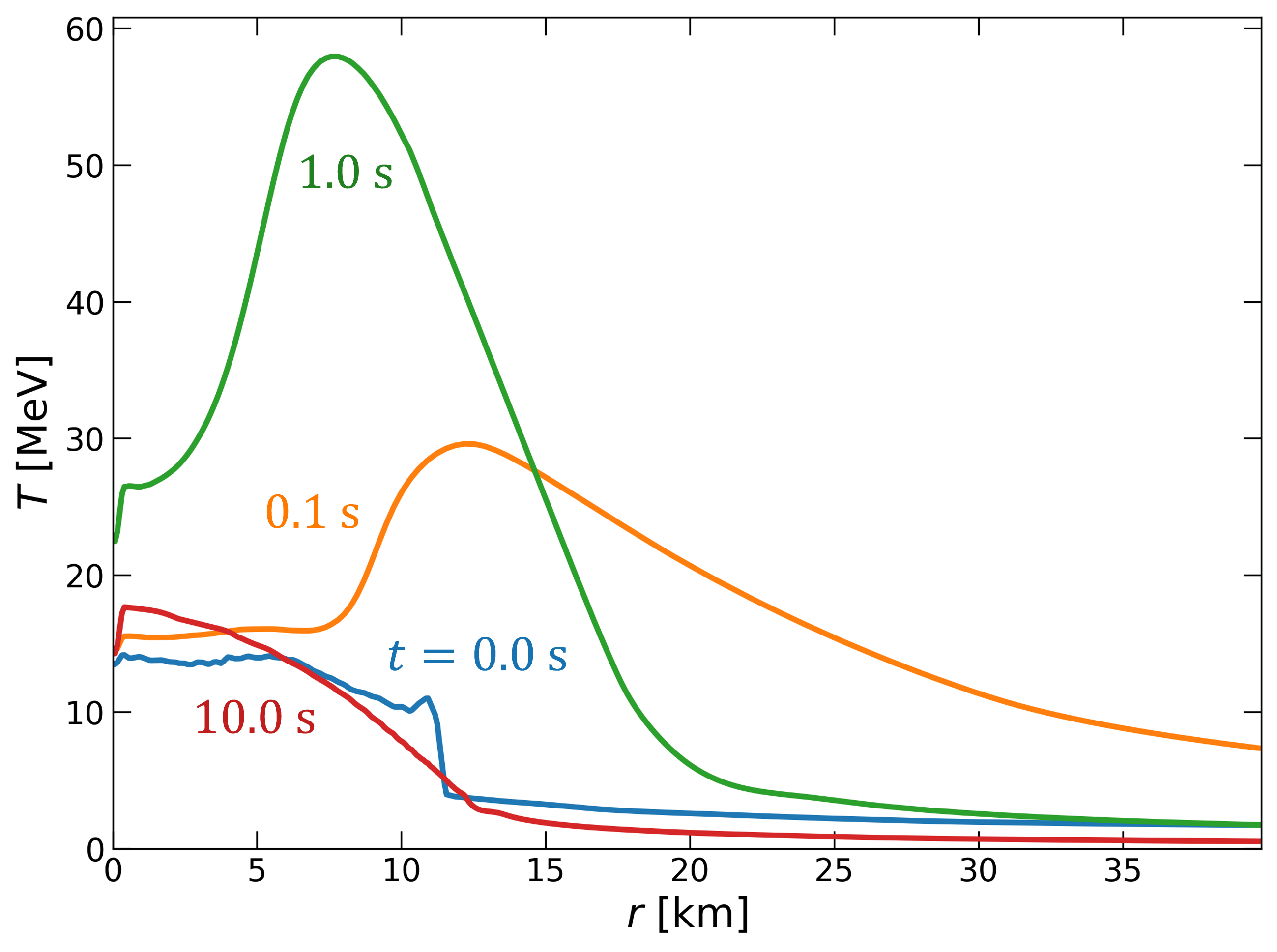}
  \captionsetup{justification=Justified}
  \caption{Snapshots of profiles of the baryon density (left panel) and temperature (right panel) as functions of distance $r$ from the supernova core at various times post core bounce obtained from the Garching 1D CCSN simulation archive~\cite{garching} with the SFHo-20 equation of state. 
  }
  \label{fig:Profile}
\end{figure*}
The temperature and density profiles of the supernova core are usually inferred from simulations. This is an active subject of research and, so far, there is no absolute convergence of the results from different simulation groups~\cite{Fiorillo:2023frv, Muller:2024slv}. These differences are mainly linked to different nuclear matter properties and equations of state, as well as different initial conditions for the proto-neutron star.    

For the results presented in this paper, we use the profiles generated by the Garching 1D CCSN model for a $20M_{\odot}$ progenitor with the SFHo equation of state  using the {\tt PROMETHEUS VERTEX} code~\cite{garching}. 
Snapshots of the supernova core density and temperature at various post-core bounce time intervals are shown in Fig.~\ref{fig:Profile} as a function of the radius. Using the available data files for various time intervals, we construct the interpolated temperature and density functions which are integrated over the burst duration and supernova volume to calculate the final emissivity.

 \subsection{Decay of Scalar}
\label{Sec:decay}
Due to its mixing with the SM Higgs boson, the scalar $S$ can decay into photons, leptons and hadrons. The partial decay widths of $S$ into all possible channels relevant in the environment of the supernova core are given in Appendix~\ref{App:Decay}   and are plotted as a function of the scalar mass in Fig.~\ref{fig:DWBR}. The photon channel is the dominant one below the electron threshold  ($m_S<2m_e\simeq 1$ MeV), whereas the $e^+e^-$ channel dominates thereafter until the muon threshold ($2m_e\leq m_S <2m_\mu\simeq 211$ MeV). The $\mu^+\mu^-$ channel briefly dominates until the pion threshold ($2m_\mu\leq m_S< 2m_\pi\simeq 270$ MeV), beyond which both muon and pion channels contribute to the decay with comparable BRs (see Fig.~\ref{fig:DWBR}). 

The decay of $S$ is relevant for our signal in two ways: \\
(i) If, after being produced inside the supernova core, it decays before escaping the core volume of radius $R_c$, it does not contribute to the emission rate in Eq.~\eqref{Eq:rate}. This is captured by the probability $P_{\rm decay}$ given by    

\begin{equation}
P_\text{decay}=\text{exp}\{-d(r,\phi)\Gamma_S\} \, ,
\label{eq:Pdecay}
\end{equation}
where $\Gamma_S\equiv \Gamma_{S}^0/\gamma_{\rm b}$ is the total decay width of $S$ in the laboratory frame, with $\gamma_{\rm b}=\omega_S/m_S$ being the Lorentz boost factor and  $\Gamma_{S}^0$ being the total proper decay width (sum of the partial widths mentioned in Appendix~\ref{App:Decay}),  and $d(r,\phi)\leq R_\nu$ is the distance $S$ travels inside the neutrinosphere of radius $R_\nu$ as a function of the radial ($r$) and angular ($\phi$) coordinates. A schematic view of the detailed geometry of $S$ traveling inside the supernova core and the form of $d(r,\phi)$ can be found in Ref.~\cite{Balaji:2022noj}. \\
(ii) Once the scalar is produced in the supernova core, to get an observable photon signal that can be clearly associated with the $S$ decay, we require that the decay must happen not just beyond $R_\nu\sim 40$ km, but beyond the supernova photosphere with radius $R_*\sim 3\times 10^{12}$ cm~\cite{Woosley:1988at,Kazanas:2014mca}. The reason is that photons will not reach the Earth if the parent $S$ decays inside the supernova photosphere, where photons are readily absorbed by the plasma. Thus, the relevant quantity for the photon signal calculation (see below) is the   
decay probability given by Eq.~\eqref{eq:Pdecay} with $d(r,\phi)$ replaced by the traversed distance $L\geq R_*$.

\begin{figure*}[t!]
  \centering
  \includegraphics[width=0.45\textwidth]{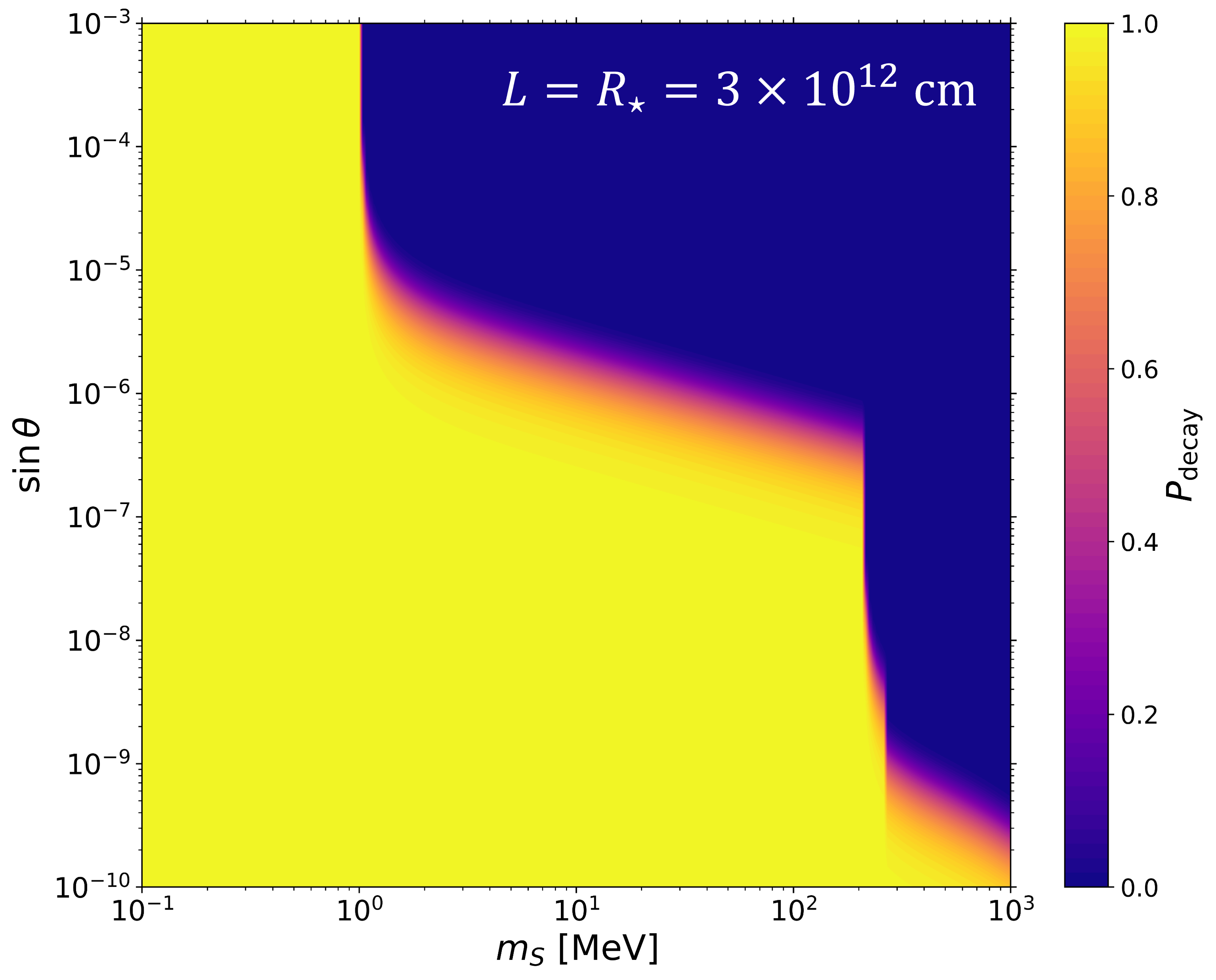}
  \hspace{0.04\textwidth}
  \includegraphics[width=0.45\textwidth]{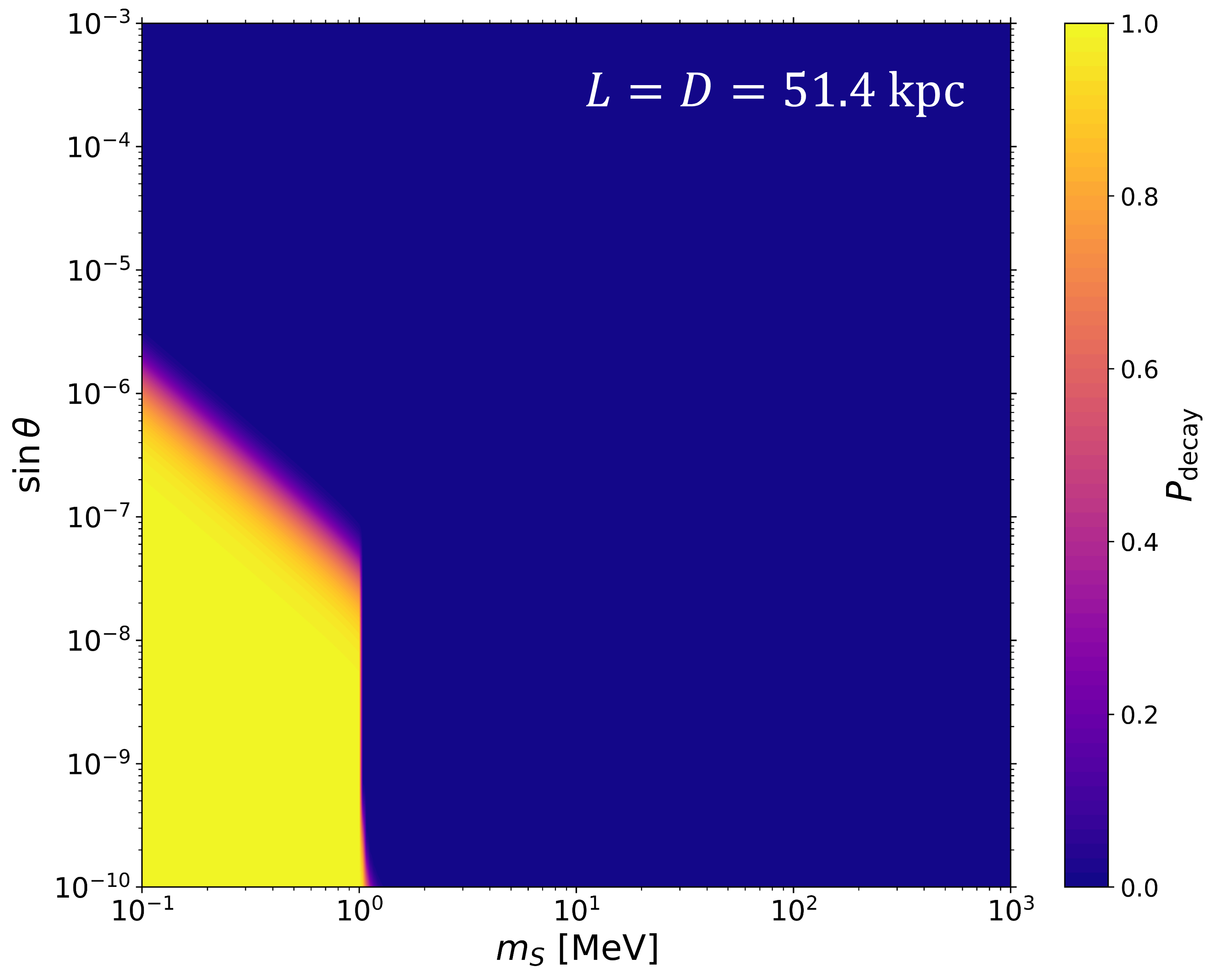}
  \captionsetup{justification=Justified}
  \caption{Heat map of the probability of scalars to decay at a  distance of $L=R_*=3\times 10^{12}$ cm (left panel) and $L=D=51.4$ kpc (right panel) after being produced in the supernova core. }
  \label{fig:SurvProb}
\end{figure*}

In Fig.~\ref{fig:SurvProb}, we show the decay probability as a function of the scalar mass and mixing angle for two benchmark distance values, namely, at $L=R_*$ (left panel) and at the Earth at a distance $D=51.4$ kpc from SN1987A~\cite{Panagia:2003rt} (right panel). As the mixing angle between the Higgs and $S$ increases, the decay of $S$ becomes more effective leading to early decay inside the supernova photosphere and thus providing no observable photon signal (blue region on the left panel, which sets the upper boundary of our signal region). The same is true for more massive scalars, as the decay rate increases with scalar mass (see Appendix~\ref{App:Decay}). On the other extreme, for very small masses and mixing angles, the scalar is too long-lived and unlikely to decay until after it has passed the Earth (blue region on the right panel, which sets the lower boundary of our signal region). The sudden drops in Fig.~\ref{fig:SurvProb} are due to various decay thresholds, namely, $e^+e^-$, $\mu^+\mu^-$, and $\pi\pi$, as we go from smaller to larger values of $m_S$.     
\subsection{Reabsorption of Scalar}
\label{Sec:reabs}
\begin{figure}[h!]
  \centering
  \includegraphics[width=0.45\textwidth]{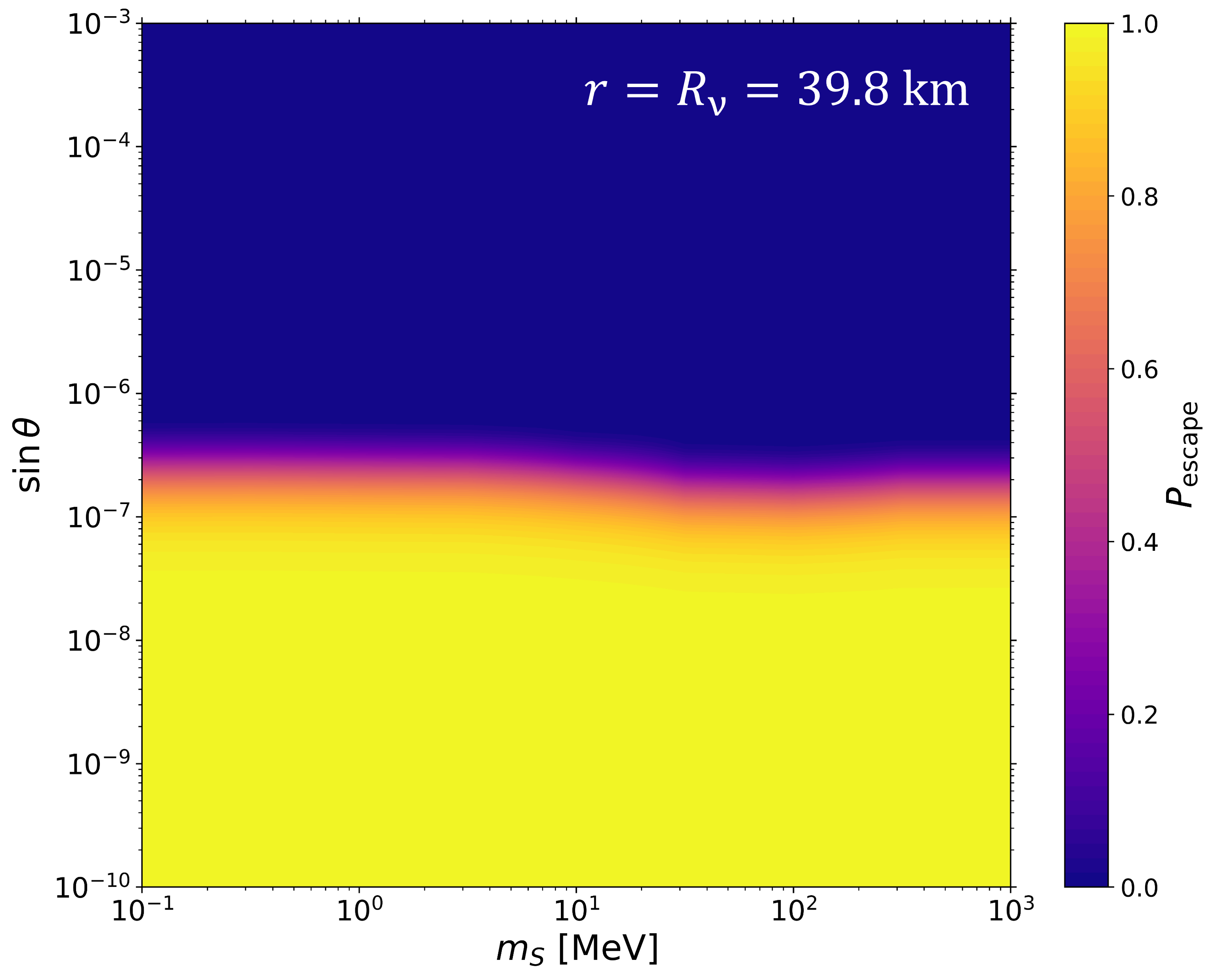}
  \captionsetup{justification=Justified}
  \caption{The heatmap of the probability of scalar not being reabsorbed within the neutrinosphere. }
  \label{fig:ReabsProb}
\end{figure}
The scalar produced inside the supernova core via bremsstrahlung process $N+N\to N+N+S$ can also be re-absorbed via inverse nucleon bremsstrahlung process $N+N+S\rightarrow N+N$, thereby  hindering them from escaping the supernova core. This is encapsulated by the probability $P_\text{escape}$ in Eq.~\eqref{Eq:rate} and is given by~\cite{Balaji:2022noj}
\begin{align}
\label{eqn:Rabs}
P_{\rm escape} \ = \ {\rm exp} \left\{
- \int_0^{d(r,\phi)} \frac{{\rm d}r'}{\lambda[L(r,\phi;r')]}\right\} \,,
\end{align}
where $\lambda$ is the MFP of the scalar inside the supernova core, which is a function of the length $L$ traversed inside the neutrinosphere. Note that $L$ is not only a function of $(r,\phi)$, but also depends on the position of $S$ along the direction of $d$ (see Fig.~3 in Ref.~\cite{Balaji:2022noj}):
\begin{align}
    L(r,\phi;r') = \sqrt{r^2+r'^2+2rr'\cos\phi} \, .
\end{align}
The inverse MFP of $S$ for the inverse bremsstrahlung process is given by~\cite{Giannotti:2005tn}
\begin{align}
    \lambda^{-1} \ = \ & \frac{1}{2\omega_S}\int {\rm d}\Pi_4 {\cal S}\sum_{\rm spins}|{\cal M}|^2\nonumber \\
    & \times (2\pi)^4\delta^4(p_1+p_2-p_3-p_4+k_S) f_1f_2 \, ,
\end{align}
where ${\rm d}\Pi_4$ is the $2\to 2$ phase space factor for the nucleon scattering. 
It is clear that the MFP depends not only on $m_S$ and $\sin\theta$ but also on the energy $\omega_S$~\cite{Dev:2020eam}. For simplicity of the calculation, an effective weighted energy-averaged MFP is used, defined as~\cite{Dev:2020eam}
\begin{eqnarray}
\langle \lambda^{-1} \rangle (m_S,\sin\theta) \ \equiv \
\frac{\int {\rm d} \omega_S \frac{\omega_S^3}{e^{\omega_S/T}-1} \lambda^{-1} (\omega_S)}{\int {\rm d} \omega_S \frac{\omega_S^3}{e^{\omega_S/T}-1}}
\,,
\end{eqnarray}
where we use the radius-dependent temperature profile.

The probability of the scalar to be reabsorbed within the neutrinosphere with radius $R_\nu=39.8$ km is shown in Fig. \ref{fig:ReabsProb}. The larger the mixing angle $\sin\theta$, the stronger is the coupling of scalars with nucleons and pions. Hence, the reabsorption probability reaches one for larger $\sin\theta$ values.

Now going back to Eq.~\eqref{Eq:rate}, we can employ some mathematical tricks to rewrite it as~\cite{Dev:2020eam}
\begin{align}
    Q &  =   \frac{n_B^2 \alpha_\pi^2 f_{pp}^4 T^{7/2} \sin^2\theta}{8 \pi^{3/2} m_N^{9/2}}\int_q^\infty {\rm d}x\int_q^\infty {\rm d}u \int_0^\infty {\rm d}v \int_{-1}^{1} {\rm d}z \nonumber \\
    & \times 
 \delta (u-v-x) \sqrt{uv} e^{-u} x \sqrt{x^2 - q^2} P_{\rm decay} P_{\rm escape}\mathcal{I}_\text{tot} \, ,
\label{Eq:EmmEnRate}
\end{align}
where  $q\equiv m_S/T$, $\alpha_\pi \equiv (2m_N/m_\pi)^2/4\pi \simeq 15$,  and $f_{pp} \simeq 1$ is the effective pion-nucleon coupling. 
The dimensionless function  $\mathcal{I}_\text{tot}$ in Eq.~\eqref{Eq:EmmEnRate} encapsulates the summed-up contributions from the $pp$, $nn$ and $np$ processes, and is given in Appendix~\ref{App:Itot}.

The number of scalars emitted per unit time per unit scalar energy can be obtained by integrating the emission rate \eqref{Eq:EmmEnRate} over the supernova volume:
\begin{align}
    \frac{{\rm d}^2N_S}{{\rm d}\omega_S {\rm d}t}&= \int {\rm d}V \frac{1}{\omega_S}\frac{{\rm d}Q}{{\rm d}\omega_S} \nonumber \\
    &= 2\pi\int_0^{R_\nu}r^2{\rm d}r\int_0^\pi \sin \phi\, {\rm d}\phi\frac{1}{\omega_S} \frac{{\rm d}Q}{{\rm d}\omega_S} .
\label{Eq:SProdRate}
\end{align}

\begin{figure}[t!]
\includegraphics[width=\linewidth]{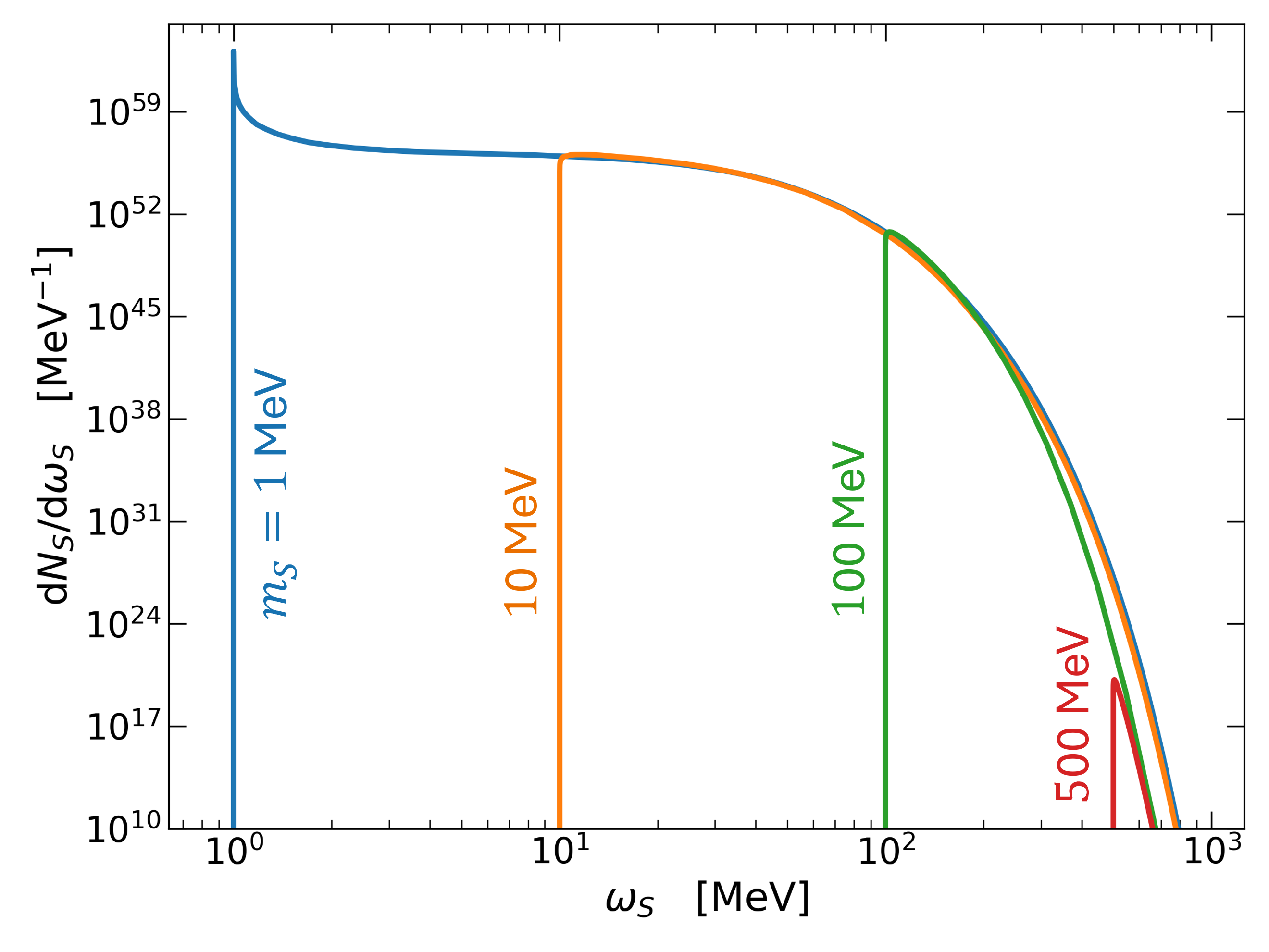} \captionsetup{justification=Justified}
    \caption{The time-integrated production rate of scalars as a function of its energy for some benchmark values of the scalar mass.}
    \label{fig:production}
\end{figure}

The time-integrated number of scalars per unit energy 
\begin{align}
    \frac{{\rm d}N_S}{{\rm d}\omega_S} \equiv \int_0^{\Delta t}dt\frac{{\rm d}^2N_S}{{\rm d}\omega_S{\rm d}t} 
\end{align}
produced in the supernova core is shown in Fig.~\ref{fig:production}. \footnote{Here we have not included the gravitational redshift and trapping effect which would essentially shift the threshold energy to $\omega_S^{\rm th}=m_S/\eta$, with $\eta=\sqrt{1-2GM/r}$ being the gravitational lapse factor~\cite{Caputo:2022mah}. This is a small effect for the supernova parameters chosen here:  $\omega_S^{\rm th}\sim 1.1m_S$~\cite{Lucente:2020whw}.} Here we have used $\Delta t=10.24$ sec for the duration of the supernova burst  and $R_\nu=39.8$ km. We find that the number of $S$ particles produced reduces significantly as the scalar mass goes beyond the mean kinetic energy of nucleons, $E_N^{\rm kin}\simeq 3.15T\sim 100$ MeV, as expected due to Boltzmann suppression. 

\section{Photon Flux from Scalar Decay}
\label{Sec:photonprod}
Following the calculation given in Refs.~\cite{Muller:2023vjm, 
Dev:2023hax}, the total photon flux coming from the various scalar decay channels takes the following general form:
\begin{align}
\label{Eq:mainprod}
&\omega_\gamma^2\dfrac{{\rm d}^2F_\gamma}{{\rm d}\omega_\gamma {\rm d}t}(\omega_\gamma,D+t) = \int_{-1}^1 {\rm d}z_\alpha\,\int_0^\infty {\rm d}L\,\dfrac{\omega_\gamma^2}{4\pi D(L_\gamma+Lz_\alpha)}\nonumber \\ 
&\quad \times \dfrac{{\rm d}^2N_\gamma}{{\rm d}\omega_\gamma {\rm d}t}(\omega_\gamma,D+t) \dfrac{m_S^2}{2\omega_S^2(1-\beta_S z_\alpha)^2}\Gamma_S\exp{(-L\Gamma_S)}\nonumber \\ 
&\quad \times \Theta(L-R_\star)\Theta(D/\sqrt{1-z_\alpha^2}-L) \,,
\end{align}
where $\rm d^2N_\gamma/(\rm d\omega_\gamma \rm dt)$ is the number of photons produced per scalar decay per unit photon energy and per unit time, $D$ is the distance between the Earth and the supernova, $L$ is the distance from the supernova at which the scalar decays, $\beta_S=\sqrt{1-m_S^2/\omega_S^2}$ is the magnitude of the velocity of the scalar, $L_\gamma$ is the distance the photon produced from the scalar decay needs to cover before reaching the earth and $z_\alpha$ is the cosine of the angle $\alpha$ that connects the direction of the scalar before its decay to the direction of the decayed photon (see Fig. \ref{fig:geometry}). The $\Gamma_S{\rm d}L\exp{(-L\Gamma_S)}$ term is the probability $P_{\rm decay}(L,L+{\rm d}L)$ of the scalar to decay within a thin shell between the distance $L$ and $L+{\rm d}L$. All the terms in Eq.~\eqref{Eq:mainprod} other than  $\rm d^2N_\gamma/(\rm d\omega_\gamma \rm dt)$ and $\Gamma_S\exp{(-L\Gamma_S)}$ come from the geometry of the scalar decay point with respect to the experimental location near Earth and the boosting of the scalar with respect to the Earth's reference frame. More details on  these terms can be found in Appendix~\ref{App:geometry}. 
The $\rm d^2N_\gamma/(\rm d\omega_\gamma \rm dt)$ term is dependent on whether the scalar decays directly to photons via $S\to \gamma\gamma$ or decays to other SM final states ($e^+e^-$, $\mu^+\mu^-$, $\pi^+\pi^-$ and $\pi^0\pi^0$ in our case) which then lead to secondary photon production via electromagnetic showers. 
In the following two subsections, we give details of the calculation of $\rm d^2N_\gamma/(\rm d\omega_\gamma \rm dt)$ for direct and secondary photon production from scalar decay.
\begin{figure*}[t!]
    \centering
   \begin{subfigure}[t]{0.48\textwidth}
       \centering
        \includegraphics[width=\linewidth]{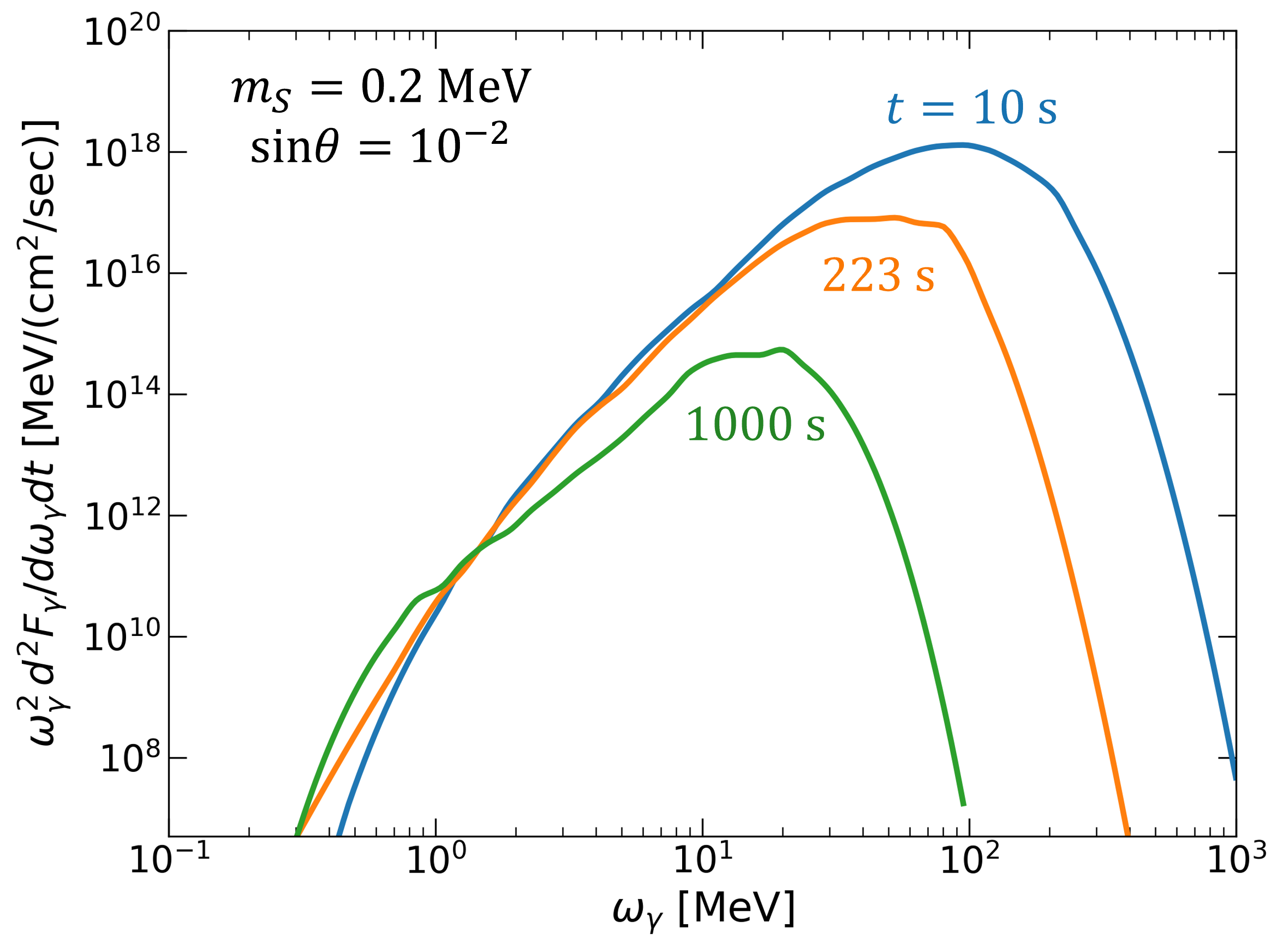}
        \label{fig:fluence_tl}
   \end{subfigure}
   \hfill
   \begin{subfigure}[t]
   {0.48\textwidth}
       \centering
        \includegraphics[width=\linewidth]{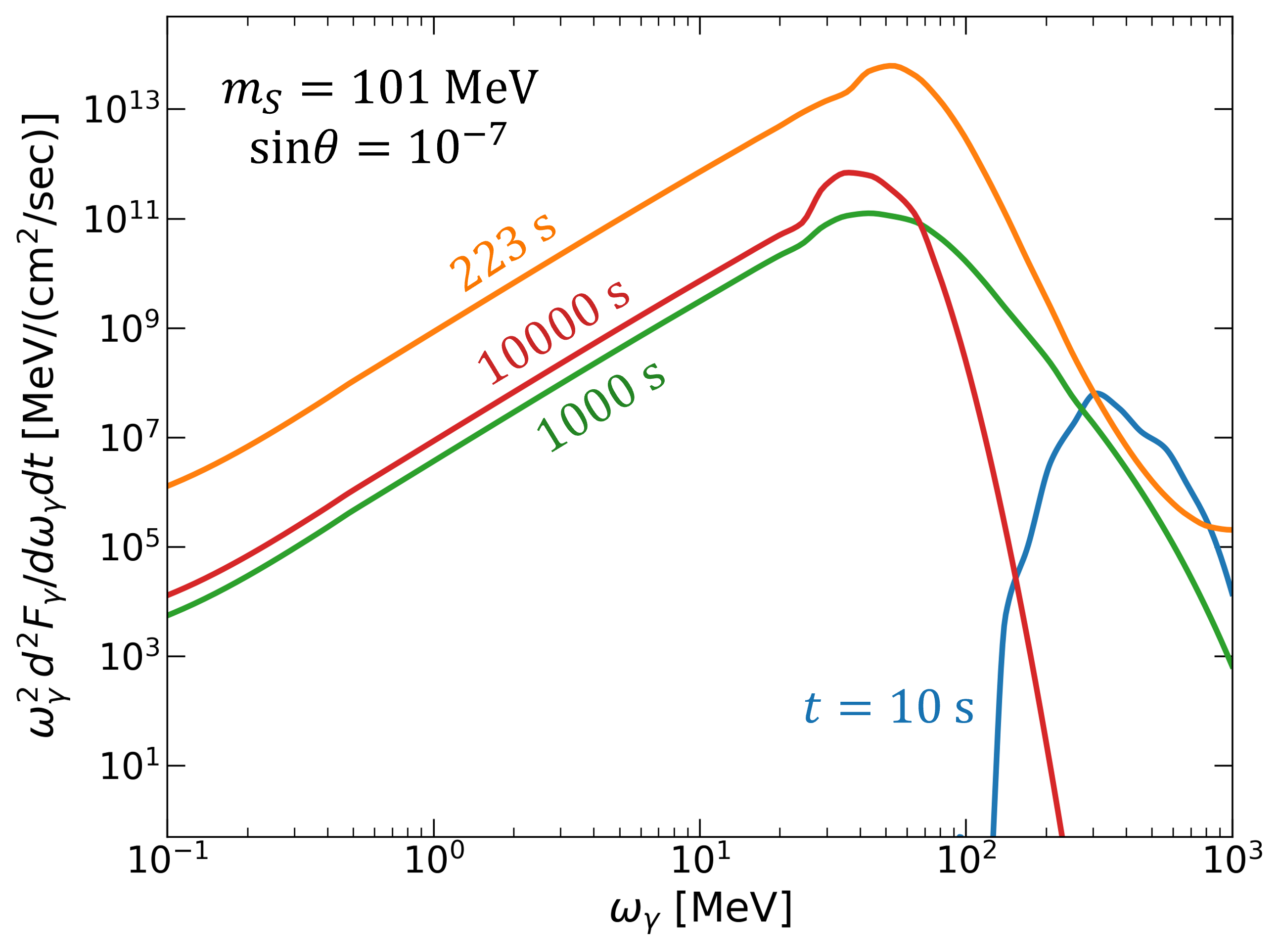}
        \label{fig:fluence_tr}
   \end{subfigure}

   \vspace{0.5em}

   \begin{subfigure}[t]{0.48\textwidth}
       \centering
       \includegraphics[width=\linewidth]{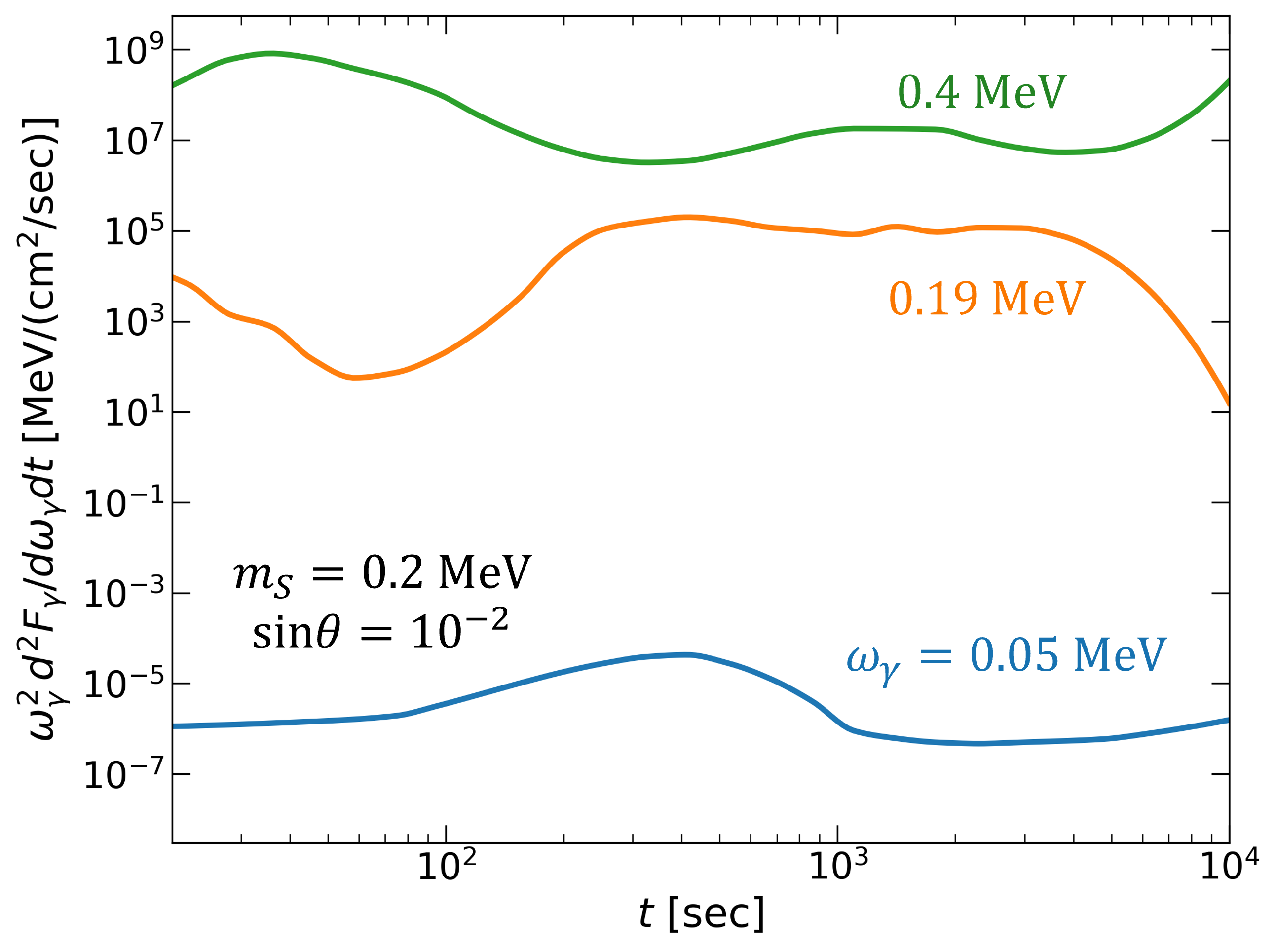}
        \label{fig:fluence_bl}
   \end{subfigure}
   \hfill
   \begin{subfigure}[t]{0.48\textwidth}
       \centering
        \includegraphics[width=\linewidth]{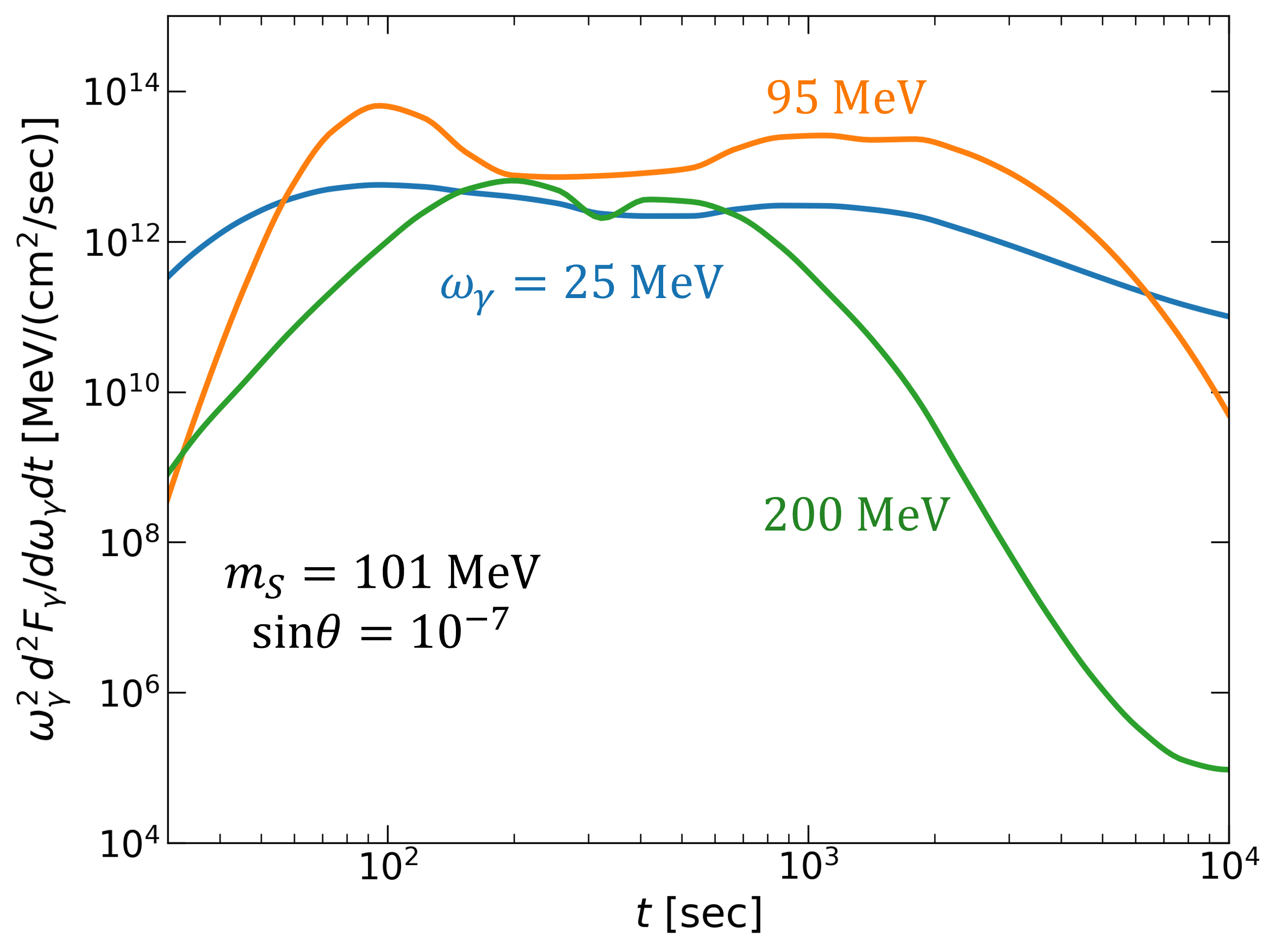}
\label{fig:fluence_br}
   \end{subfigure}
\captionsetup{justification=Justified}
    \caption{Differential photon fluence as a function of energy (top row) and as a function of time (bottom row) for two different benchmark values of $(m_S,\sin\theta)=(0.2~{\rm MeV},10^{-2})$ (left panel) and $(101~{\rm MeV},10^{-7})$ (right panel).}
    \label{fig:Fluence}
\end{figure*}
\subsection{Direct Photon Production}
The number of direct photons from the decay of a single scalar per unit photon energy and per unit time is given by 
\begin{align}
    \dfrac{{\rm d}^2N_\gamma}{{\rm d}\omega_\gamma {\rm d}t}(\omega_\gamma,D+t)\ = \ & 2\frac{{\rm d}^2N_S}{{\rm d}\omega_S {\rm d}t}(\omega_\gamma,D+t-L/\beta_S-L_\gamma)\nonumber \\ 
& \qquad \times \text{Jac}(\omega_S,\omega_\gamma)
\label{Eq:DirectNgamma}
\end{align}
where $\text{Jac}(\omega_S,\omega_\gamma)= |\partial\omega_S/\partial\omega_\gamma|$ is the Jacobian and the scalar energy and photon energy are related by 
\begin{equation}   \omega_S=\dfrac{m_S^2\pm\sqrt{z_\alpha^2m_S^2[m_S^2-4(1-z_\alpha^2)\omega_\gamma^2]}}{2(1-z_\alpha^2)\omega_\gamma} \, ,
\end{equation}
where the $\pm$ sign takes into account the fact that there are two possible triangles associated with the geometry of the decay\cite{Dev:2023hax}. 
The factor $2$ in Eq.~\eqref{Eq:DirectNgamma} comes from the production of two photons per single scalar decay.
\subsection{Secondary Photon Production}
The number of secondary photons from the decay of a single scalar per unit photon energy and per unit time are given by :
\begin{align}
    \dfrac{{\rm d}^2N_\gamma}{{\rm d}\omega_\gamma {\rm d}t}(\omega_\gamma,D+t)& =\int {\rm d}\omega_S \frac{{\rm d}^2N_S}{{\rm d}\omega_S {\rm d}t}(D+t-L/\beta_S-L_\gamma)\nonumber \\
    & \qquad \times \left.\frac{{\rm d}N_\gamma}{{\rm d}\omega_\gamma}\right|_\text{Boosted}(\omega_S,\omega_\gamma) \, ,
\label{Eq:IndirectNgamma}
\end{align}
where ${\rm d}N_\gamma/{\rm d}\omega_\gamma$ is the secondary photon spectrum from the decay of a single scalar of a particular energy $\omega_S$. This is calculated for the $e^+e^-$ and $\mu^-\mu^-$ channels using an analytic form given in Ref.~\cite{Fortin:2009rq} and also numerically using {\tt PYTHIA}~\cite{Sjostrand:2014zea}, both of which match quite well; see Appendix~\ref{App:pythiamatch} for details. For computational simplicity, we do not include the hadronic channels here, which could in principle enhance the photon flux beyond the two-pion threshold, but will have a negligible effect on the overall result, since the scalar emission is Boltzmann-suppressed in this high-mass regime (cf.~Fig.~\ref{fig:production}).  

\subsection{Total Differential Photon Fluence}
The total differential photon fluence from the decay of scalar via photon, electron and muon channels is given by 

\begin{align}
    \omega_\gamma^2\dfrac{{\rm d}^2F_\gamma}{{\rm d}\omega_\gamma {\rm d}t}\Big|_\text{tot}=& {\rm BR}(S\to \gamma\gamma)\,\omega_\gamma^2\dfrac{{\rm d}^2F_\gamma}{{\rm d}\omega_\gamma {\rm d}t}\Big|_{\gamma\gamma}\nonumber \\
    & +{\rm BR}(S\to e^+e^-)\,\omega_\gamma^2\dfrac{{\rm d}^2F_\gamma}{{\rm d}\omega_\gamma {\rm d}t}\Big|_{e^+e^-}\nonumber \\
    &+{\rm BR}(S\to \mu^+\mu^-)\, \omega_\gamma^2\dfrac{{\rm d}^2F_\gamma}{{\rm d}\omega_\gamma {\rm d}t}\Big|_{\mu^+\mu^-} \, ,
\end{align}
where the individual fluences in each channel is calculated using the master formula~\eqref{Eq:mainprod}, with the ${\rm d}^2N_\gamma/({\rm d}\omega_\gamma{\rm d}t)$ part for the $\gamma\gamma$ channel given by Eq.~\eqref{Eq:DirectNgamma} and for the lepton channels given by Eq.~\eqref{Eq:IndirectNgamma}, respectively.

In Fig.~\ref{fig:Fluence}, we show the total differential photon fluence. The top row shows the fluence as a function of photon energy at different post-core bounce times, whereas the bottom row shows the fluence as a function of time for different photon energies. The left panels correspond to a lighter scalar with larger mixing angle, whereas the right panel is for a heavier scalar with smaller mixing angle. These benchmark points are chosen to ensure that the scalar decay probability is large (cf.~Fig.~\ref{fig:SurvProb} left panel). Although large values of $\sin\theta$ result in a highly suppressed escape probability due to small MFP (cf.~Fig.~\ref{fig:ReabsProb}), it is still relevant for the photon signal, as we will see in the next section.      
\section{Results}
\label{Sec:results}
For the photon signal from SN1987A to be observable, the scalars have to be produced during the supernova-burst and the produced photon signal has to reach the GRS on board SMM within its observation time window of $t_{\rm M}=223$ sec from the onset of the supernova burst ($t=0$) as determined by the first neutrino observation. Hence, we rewrite the differential production rate of the scalars in the supernova [cf.~Eqs.~\eqref{Eq:DirectNgamma} and \eqref{Eq:IndirectNgamma}] in the following way:
\begin{equation}
\dfrac{{\rm d}^2N_S}{{\rm d}\omega_S {\rm d}t}(\omega_S,t_S)=\dfrac{{\rm d}^2N_S}{{\rm d}\omega_S {\rm d}t}(\omega_S)\Theta(t_S)\Theta(\Delta t-t_S) \,,
\end{equation}
where $t_S=t+D-L/\beta_S-L_\gamma$ is the time at which scalar is produced and $\Delta t=10.24$ is the duration of the supernova burst.   
Thus, we can calculate total photon fluence observed in the instrument time and energy window using the following relation:
\begin{equation}
F_\gamma(m_S,\sin\theta)=\int_{\omega_1}^{\omega_2}{\rm d}\omega_\gamma \int_0^{t_\text{M}} {\rm d}t \frac{{\rm d}^2F}{{\rm d}\omega_\gamma {\rm d}t}(m_S,\sin\theta,\omega_\gamma,D+t) \, ,
\end{equation}
where $\omega_1=25$ MeV and $\omega_2=100$ MeV are the minimum and maximum energy of photons recorded by SMM in the observation window of $t_{\rm M}=223$ sec~\cite{Chupp:1989kx}.

\begin{figure}[t!]
    \centering
    \includegraphics[width=0.48\textwidth]{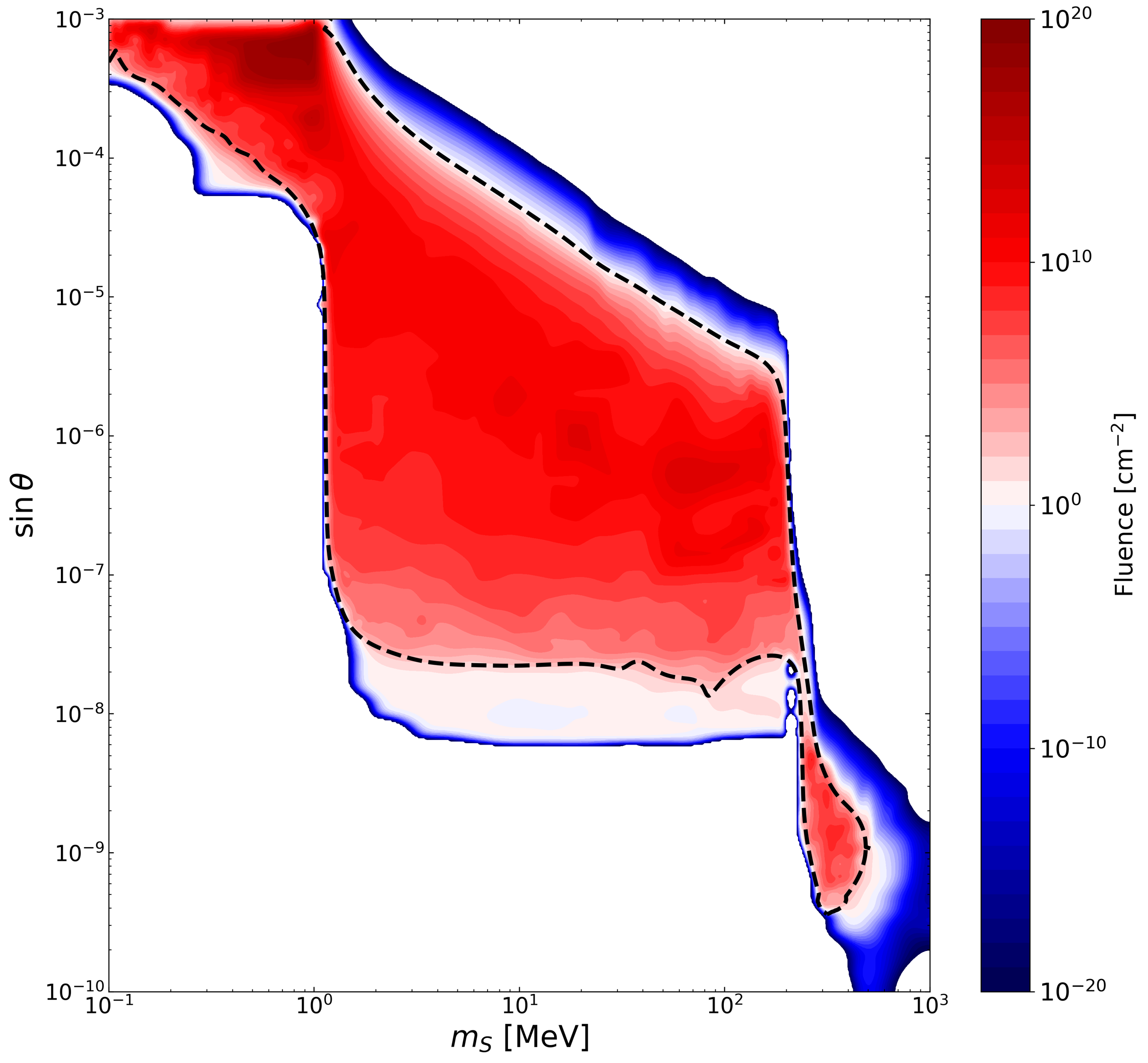}
    \captionsetup{justification=Justified}
    \caption{Heat map of the total photon fluence from the decay of scalars produced in SN1987A core in the $(m_S,\sin\theta)$ plane. The dashed contour shows the SMM upper limit on the fluence, which gives us the new gamma-ray limit shown in Fig.~\ref{fig:Constraint}.}
    \label{fig:Contour}
\end{figure}

The heatmap of the photon fluence over the ($m_S,\sin\theta)$ parameter space is shown in Fig.~\ref{fig:Contour}. The contour plot shows three distinct regions, as in Fig.~\ref{fig:SurvProb}-- the first region extends up to $m_S\sim 2m_e$, the second one extends till $m_S\sim 2m_\mu$ and the third region corresponds to $m_S$ beyond $2m_\mu$. The first region indicates the contribution of direct photon decay of scalar in the total photon fluence. This region is rather narrow, only appearing at higher values of $\sin\theta$ mainly because of the loop-suppressed $S\to \gamma\gamma$ decay rate. When the scalar mass reaches $2m_e$, the $e^+e^-$ decay channel opens up leading to a broader region of $(m_S,\sin\theta)$ parameter space being excluded. It is important to note that for $m_S\gtrsim2m_e$, though the direct photon decay channel is subdominant compared to the $e^+e^-$, due to more photons being produced from the direct decay of scalar into photons, the contribution of direct photon fluence in the total photon fluence from the scalar decay is still quite significant. The direct photon decay channel becomes subdominant for $m_S\gtrsim2m_\mu$ (see Fig.~\ref{fig:DWBR}) and hence, the third region in the contour plot in Fig.~\ref{fig:Contour} arises mainly from the secondary photons produced from the $\mu^+\mu^-$ decay channel of $S$.  As $\sin\theta$ increases, the MFP of the scalars drops and eventually the Stefan-Boltzmann limit (where the emission comes solely from the surface) is approached \cite{Burrows:1990pk,Caputo:2021rux}.  We find that this emission is sufficient to produce enough photons to constrain some of this high-mixing area of parameter space (the neck region in Fig.~\ref{fig:Contour}).

\begin{figure*}[t!]
    \centering \includegraphics[width=1.0\textwidth]{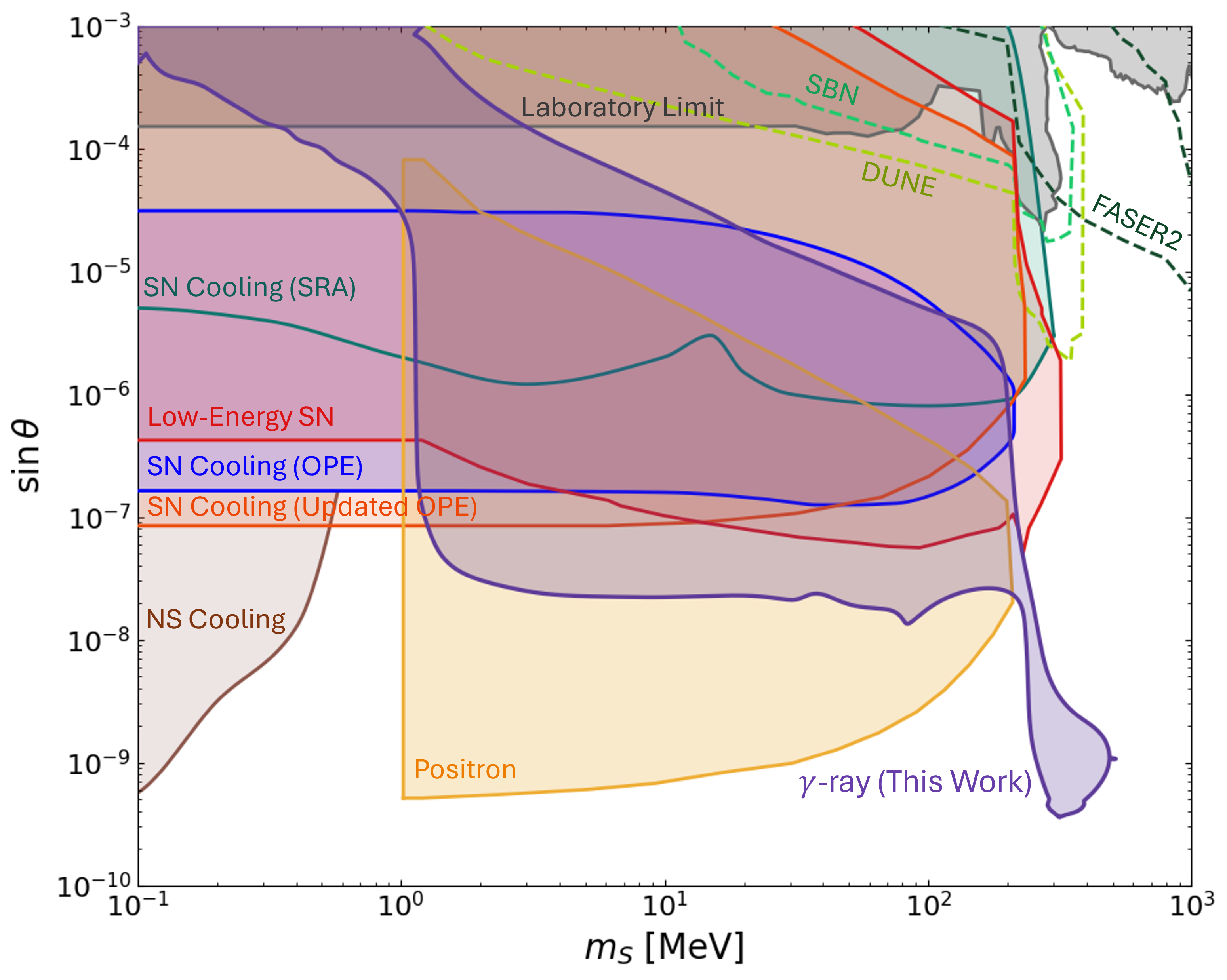}
    \captionsetup{justification=Justified}
    \caption{ The new gamma-ray constraint from SN1987A (purple-shaded region) in the scalar mass-mixing plane. For comparison, we also show the SN1987A cooling bounds derived using OPE~\cite{Balaji:2022noj} (blue-shaded) and SRA~\cite{Hardy:2024gwy} (dark-green-shaded) approximations. Improved SN bounds using an updated OPE calculation of SN1987A (orange-shaded) and low-energy SN cooling (red-shaded) bounds, as well as positron emission (yellow-shaded)~\cite{Joseph:2026nut} are also shown. In addition, the gray-shaded region on top is excluded from various laboratory searches, and the dashed contours show the projected sensitivity of future laboratory experiments~\cite{Batell:2022dpx, Feng:2022inv}. }
    \label{fig:Constraint}
\end{figure*}

SMM did not register any significant photon signal during its operation and thus set an upper limit of total fluence to be $1.78$ cm$^{-2}$~\cite{Chupp:1989kx}. Using this upper limit on total photon fluence, we obtain the constraint on $(m_S,\sin\theta)$ parameter space which is shown as the dashed black contour in Fig.~\ref{fig:Contour} and as the purple-shaded region in Fig.~\ref{fig:Constraint} along with other existing constraints for comparison. 

The blue shaded region in Fig.~\ref{fig:Constraint} is the excluded parameter space from the cooling of SN1987A using the OPE scheme to calculate the scalar production in supernova~\cite{Balaji:2022noj}. This constraint was recently reevaluated using an updated OPE prescription of scalar production in supernova~\cite{Joseph:2026nut} and the corresponding exclusion region is shown by the orange-shaded region in Fig.~\ref{fig:Constraint}. The teal-shaded region in Fig.~\ref{fig:Constraint} represents the excluded parameter space from SN1987a cooling, where scalar production from resonant mixing with the in-medium longitudinal photon is considered, which dominates for $m_S\lesssim 10\text{ MeV}$, and production from nucleon bremsstrahlung, which dominates for $m_S\gtrsim 10\text{ MeV}$, is calculated within the framework of the soft radiation approximation (SRA)\footnote{The SRA allows one to relate the scalar production rate, in the soft limit, to on-shell nucleon-nucleon scattering data \cite{Hanhart:2000ae}.  However, high-mass scalars must necessarily be produced with energy $>m_S$, and thus the SRA becomes increasingly worse for high-mass scalars.  For a more extensive discussion, see the review~\cite{Fortin:2021cog}.}. The gray shaded parameter space in Fig.~\ref{fig:Constraint} denotes the excluded region from various laboratory searches and the dashed lines correspond to the sensitivity of future experiments in the $m_S-\sin\theta$ parameter space~\cite{Batell:2022dpx, Feng:2022inv}. At lower masses (below MeV), there exist other astrophysical constraints from neutron star cooling~\cite{Fiorillo:2025zzx} (brown-shaded);  other stellar cooling bounds~\cite{Dev:2020jkh,Balaji:2022noj,Bottaro:2023gep, Yamamoto:2023zlu} do not show up in the mass range shown here.  Similarly, we do not show any cosmological bound~\cite{Fradette:2018hhl} (from BBN and CMB) in Fig.~\ref{fig:Constraint} because the scalars do not come to equilibrium at BBN temperature for the parameter space shown here.   

Following Refs.~\cite{Diamond:2021ekg,Diamond:2023scc, Diamond:2023cto}, we have also checked that in the region of our interest newly constrained by the gamma-ray limit, there is no fireball formation, which would have otherwise  depleted the gamma-ray photons into lower energies. 

Comparing our new gamma-ray constraint with the recently calculated positron constraint (yellow-shaded) in Fig.~\ref{fig:Constraint}, we find that the secondary gamma-ray photons from the scalar decay to $\mu^+\mu^-$ rule out new parameter space  around $\sin\theta=10^{-9}-10^{-8}$ for $m_S\gtrsim 210$ MeV.

\section{Conclusion}
\label{Sec:conclusion}

In this work, we have derived new constraints on a light CP-even scalar coupled to the SM Higgs by revisiting archival gamma-ray data from SN1987A. The central new ingredient in our analysis is the inclusion of secondary photon production from the decay of the scalar into charged leptons. Indeed, once the decay channels $S\to e^+e^-$ and $S\to \mu^+\mu^-$ open up, the additional photon contribution  through secondary emission from its charged decay products must be added to the direct decay mode $S\to\gamma\gamma$. Incorporating this effect into the total photon fluence is one of the main new results of our analysis, and it substantially changes the interpretation of the SN1987A gamma-ray bound.

The important point is that in setting bounds on CP-even scalars from SN1987A, the problem should be treated as an inherently multichannel one. After being produced through nucleon bremsstrahlung, the scalar propagates outwards through the supernova while being attenuated by decay and reabsorption by inverse bremsstrahlung; ultimately, it contributes to the photon signal through several decay pathways, of which direct decay to photons is only one.  At low $S$ masses, the signal is controlled primarily by the direct  two photon channel. Once the electron threshold is crossed, however, the decay $S\to e^+e^-$ opens up and enlarges the excluded region through the associated secondary photon contribution. At still higher masses, the opening of the muon channel further reshapes the bound, and for sufficiently large $m_S$ the photon signal is dominated by secondary photons from $S\to \mu^+\mu^-$.

By comparing this total predicted fluence with the negative observation of an excess by the Solar Maximum Mission in the time window following the SN1987A neutrino burst, we obtain a new exclusion region in the $(m_S,\sin\theta)$ plane. Our final results are depicted in Fig.~\ref{fig:Constraint}. Our constraint is complementary to supernova cooling, cosmological, and laboratory bounds, and is especially relevant in the regime where the scalar is weakly enough coupled to escape the core yet sufficiently short lived to decay into visible final states on astrophysically relevant length scales. It thus probes parameter space that is not captured by cooling arguments alone.

\acknowledgments
The work of B.D.~and W.M.~was partly supported by the US Department of Energy under grant No. DE-
SC0017987.  J.F.F.~is supported by NSERC.  S.P.H.~acknowledges the support of the National Science Foundation grant PHY 21-16686.  K.S.~is supported in part by the National Science Foundation under Award No. PHY-2514896. YZ is supported by the National Natural Science Foundation of China under grant No.~12175039 and the ``Fundamental Research Funds for the Central Universities".

\medskip 
\noindent
{\bf Note Added:} Preliminary results of our analysis were presented at the PPP conference in November 2025~\cite{PPP}.  While finalizing our paper for arXiv submission, Ref.~\cite{Joseph:2026nut} appeared, which presents improved supernova bounds on CP-even scalars. While the basic formalism is similar to ours, we notice the following differences: (i) They do not consider the decay channel $S\to \gamma\gamma$ (either primary or secondary photons), which is our main focus here.  (ii) They do not include the angular dependence of the mean free path. (iii) They have used a different effective coupling of $S$ to pions. A preliminary comparison of our gamma-ray constraints with their cooling and positron constraints is shown in Fig.~\ref{fig:Constraint}.      

\appendix
\onecolumngrid
\section{Partial Decay Widths of Scalar $S$}
\label{App:Decay}
\begin{figure*}[t!]
  \centering
  \includegraphics[width=0.45\textwidth]{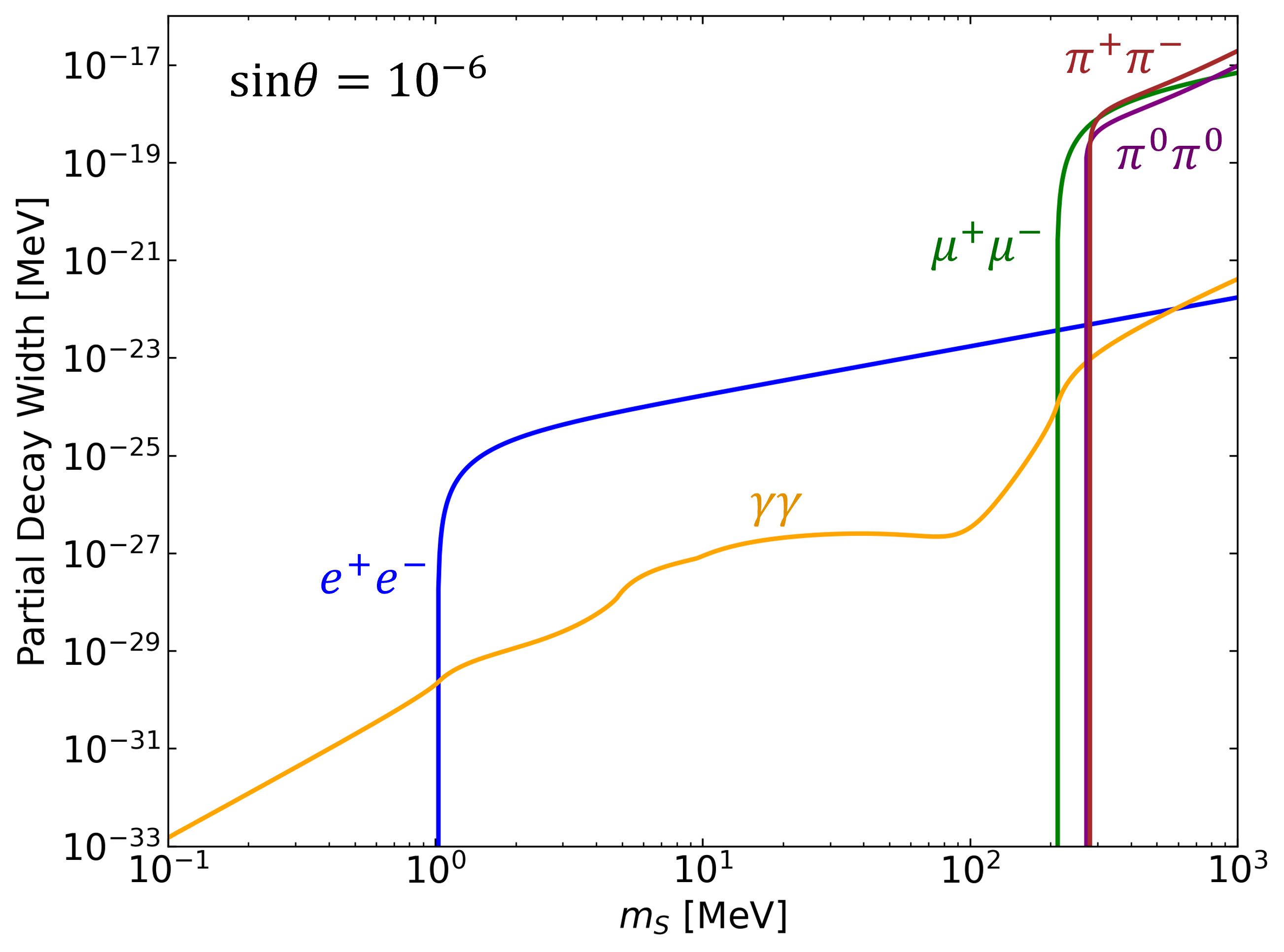}
  \hspace{0.04\textwidth}
  \includegraphics[width=0.45\textwidth]{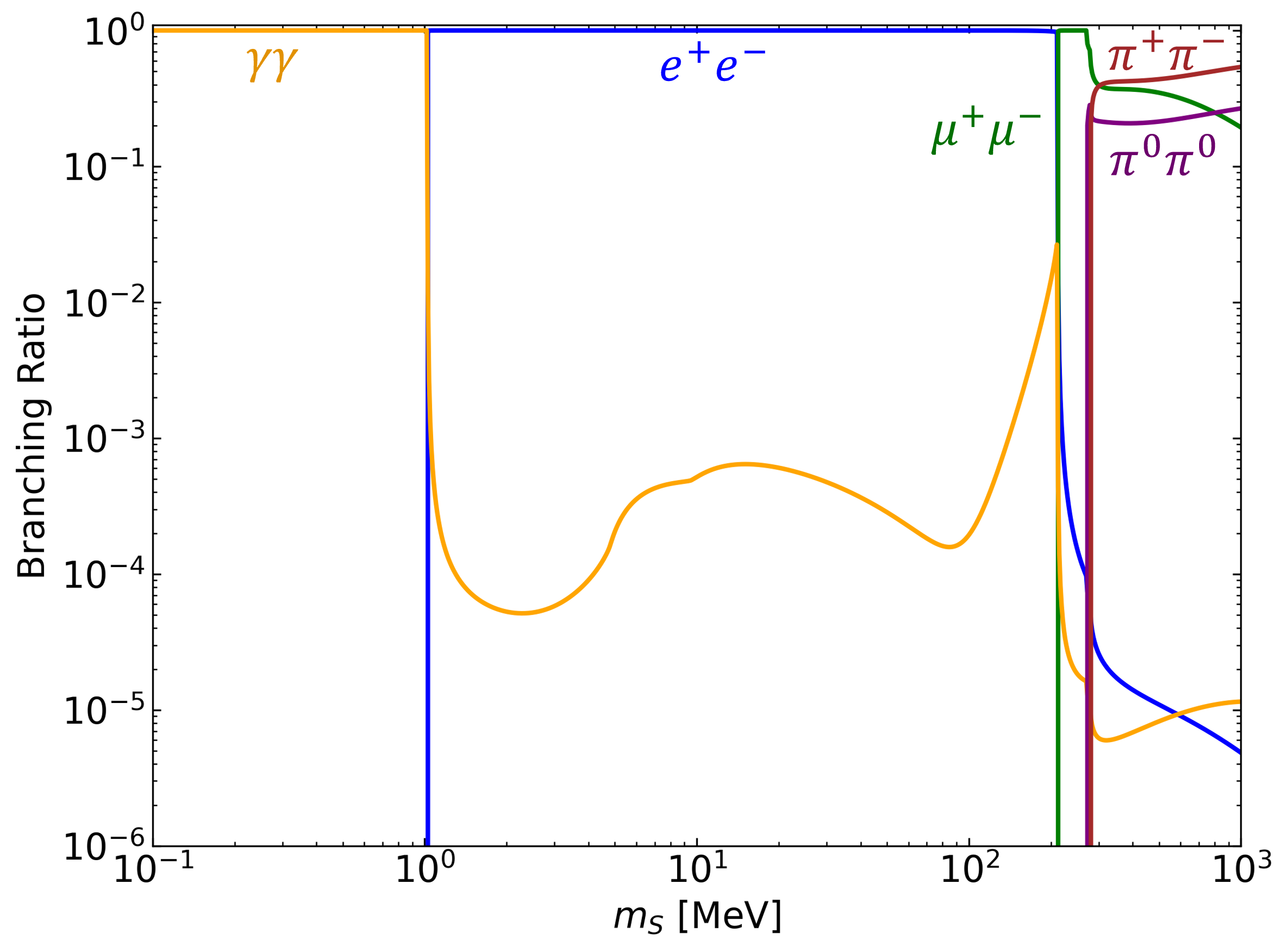}
  \captionsetup{justification=Justified}
  \caption{{\it Left:} The partial decay widths of the scalar to various SM final states as a function of the scalar mass for $\sin\theta=10^{-6}$. For other values of $\sin\theta$, the decay widths can simply be scaled as $\sin^2\theta$. {\it Right:} The decay branching ratios, which are independent of $\sin\theta$.}
  \label{fig:DWBR}
\end{figure*}

The proper partial decay widths of the scalar for the relevant channels in the supernova environment are written below~\cite{Dev:2020eam}:
\begin{eqnarray}
\Gamma_0 (S \to e^+e^-) & \ = \ &
\frac{m_S m_e^2 \sin^2\theta}{8\pi v_{\rm EW}^2}
\left( 1-\frac{4m_e^2}{m^2_{S} }\right)^{3/2} \,, \\
\Gamma_0 (S \to \mu^+\mu^-) & \ = \ &
\frac{m_S m_\mu^2 \sin^2\theta}{8\pi v_{\rm EW}^2}
\left( 1-\frac{4m_\mu^2}{m^2_{S} }\right)^{3/2} \,, \\
\label{eqn:Saa}
\Gamma_0 (S \to \gamma\gamma) & \ = \ &
\frac{\alpha^2 m_{S}^3 \sin^2\theta}{256\pi^3 v_{\rm EW}^2}
\left| \sum_f N_C^f Q_f^2 A_{1/2} (\tau_f) + A_1 (\tau_W) \right|^2 \,, \\
\Gamma_0 (S \to \pi^0 \pi^0) & \ = \ &
\frac{\sin^2\theta}{648\pi m_S v_{\rm EW}^2}
\left( m_S^2+ \frac{11}{2} m_{\pi^0}^2 \right)^2
\left( 1-\frac{4m_{\pi^0}^2}{m_S^2} \right)^{1/2} \,,  \\
\Gamma_0 (S \to \pi^+ \pi^-) & \ = \ &
\frac{\sin^2\theta}{324\pi m_S v_{\rm EW}^2}
\left( m_S^2+ \frac{11}{2} m_{\pi^\pm}^2 \right)^2
\left( 1-\frac{4m_{\pi^\pm}^2}{m_S^2} \right)^{1/2} \,,
\end{eqnarray}

where $m_e,\,m_\mu,\, m_{\pi^0},\, m_{\pi^\pm}$ are the masses of $e$, $\mu$, $\pi^0$ and $\pi^\pm$ respectively; $\alpha=e^2/4\pi$ is the fine-structure constant; $N_C=3~(1)$ is the color factor for quarks (charged leptons) and $A_{1/2}(\tau_f),~A_1(\tau_W)$ (with $\tau_X=m_S/4m_X^2$) are the loop functions that can be found in Appendix A of Ref.~\cite{Dev:2017dui}.  These partial decay widths and the corresponding branching ratios (BRs) as a function of the scalar mass are shown in Fig.~\ref{fig:DWBR}. The BRs obtained here agree with  Ref.~\cite{Winkler:2018qyg} and with Figure 4.15 lower panel in Ref.~\cite{Feng:2022inv}. 
 
\section{Scalar Emission Rate}
\label{App:Itot}
The dimensionless function ${\cal I}_{\rm tot}$ in Eq.~(\ref{Eq:EmmEnRate}) is given by~\cite{Balaji:2022noj}\footnote{This expression differs from that in Ref.~\cite{Dev:2020eam} by a factor of 1/4 and 1/2 for the first and third terms, respectively, which is due to the higher-order corrections in expansion of the small parameters~\cite{Dev:2021kje}.}
\begin{eqnarray}
\label{eqn:Itot}
{\cal I}_{\rm tot} \ = \
\frac14 y_{hNN}^2 \left( \frac{q}{x} \right)^4 {\cal I}_A +
\frac{1}{81} \left( \frac{m_N}{v_{\rm EW}} \right)^2 {\cal I}_B +
\frac{1}{9} y_{hNN} \left( \frac{q}{x} \right)^2
\left( \frac{m_N}{v_{\rm EW}} \right) {\cal I}_C \,.
\end{eqnarray}
Each of ${\cal I}_{A,\, B,\, C}$ is a sum of the $pp$, $nn$ and $np$ contributions: 
\begin{eqnarray}
\label{eqn:Iabc}
{\cal I}_{A,\, B,\, C} \ = \ {\cal I}_{A,\, B,\, C}^{(pp)} + {\cal I}_{A,\, B,\, C}^{(nn)} + 4{\cal I}_{A,\, B,\, C}^{(np)} \,,
\end{eqnarray}
with ${\cal I}_{A,\, B,\, C}^{(pp)} = {\cal I}_{A,\, B,\, C}^{(nn)}$, and 
\begin{eqnarray}
\label{eqn:IApp}
{\cal I}_{A}^{(pp)} & \ = \ &  
\frac{c_k^2}{c_{k\pi}^2} + \frac{c_l^2}{c_{l\pi}^2} - \frac{c_{kl}^2}{c_{k\pi}c_{l\pi}} \,, \\
\label{eqn:IBpp}
{\cal I}_{B}^{(pp)} (r) & \ = \ &  
\left( q^2 \frac{T(r)}{m_N} + \frac{11}{2} y \right)^2
\left[ \frac{c_k^2}{c_{k\pi}^4} + \frac{c_l^2}{c_{l\pi}^4} - \frac{c_{kl}^2}{c_{k\pi}^2c_{l\pi}^2} \right] \,, \\
\label{eqn:ICpp}
{\cal I}_{C}^{(pp)} (r) & \ = \ & 
\left( q^2 \frac{T(r)}{m_N} + \frac{11}{2} y \right)
\left[ \frac{c_k^2}{c_{k\pi}^3} + \frac{c_l^2}{c_{l\pi}^3} - \frac{c_{kl}^2}{2c_{k\pi}c_{l\pi}^2} - \frac{c_{kl}^2}{2c_{k\pi}^2c_{l\pi}} \right] \,, \\
\label{eqn:IAnp}
{\cal I}_{A}^{(np)} & \ = \ &
\frac{c_k^2}{c_{k\pi}^2} + \frac{4c_l^2}{c_{l\pi}^2} + \frac{2c_{kl}^2}{c_{k\pi}c_{l\pi}} \,, \\
\label{eqn:IBnp}
{\cal I}_{B}^{(np)} (r) & \ = \ &
\left( q^2 \frac{T(r)}{m_N} + \frac{11}{2} y \right)^2
\left[ \frac{c_k^2}{c_{k\pi}^4} + \frac{4c_l^2}{c_{l\pi}^4} + \frac{2c_{kl}^2}{c_{k\pi}^2c_{l\pi}^2} \right] \,, \\
\label{eqn:ICnp}
{\cal I}_{C}^{(np)} (r) & \ = \ &
\left( q^2 \frac{T(r)}{m_N} + \frac{11}{2} y \right)
\left[ \frac{c_k^2}{c_{k\pi}^3} + \frac{4c_l^2}{c_{l\pi}^3} + \frac{c_{kl}^2}{c_{k\pi}c_{l\pi}^2} + \frac{c_{kl}^2}{c_{k\pi}^2c_{l\pi}} \right] \,.
\end{eqnarray}
The $c$ functions are defined as
\begin{eqnarray}
c_k & \ \equiv \ &  u+v-2z\sqrt{uv}  \,, \\
c_{k\pi} & \ \equiv \ &  u+v+y-2z\sqrt{uv} \,, \\
c_l & \ \equiv \ &  u+v+2z\sqrt{uv} \,, \\
c_{l\pi} & \ \equiv \ &  u+v+y+2z\sqrt{uv} \,, \\
c_{kl}^2 & \ \equiv \ &
 u^2 + v^2 + 2 u v (-3 + 2 z^2) \,.
\end{eqnarray}
The dimensionless functions are defined as~\cite{Giannotti:2005tn, Dent:2012mx}
\begin{align}
\label{eq:uvxyqz}
& u \ \equiv \ \frac{{\bf p}_i^2}{m_N T} \,, \quad
 v \ \equiv \ \frac{{\bf p}_f^2}{m_N T} \,, \quad
 x \ \equiv \ \frac{E_S}{T} \,,  \quad
 q \ \equiv \ \frac{m_S}{T} \,, \quad
 y \ \equiv \ \frac{m_{\pi}^2}{m_N T} \, , \quad
r \ \equiv \ \frac{T}{m_N} \,, \quad
 z \ \equiv \ \cos(\theta_{if}) \,,
\end{align}
where ${\bf p}_i$ and ${\bf p}_f$ are the three-momenta of the initial and final-state nucleons and $\theta_{if}$ is the scattering angle. 
\section{Geometry of Scalar Decay}
\label{App:geometry}
\begin{figure}[t]
    \centering
    \includegraphics[width=0.5\linewidth]{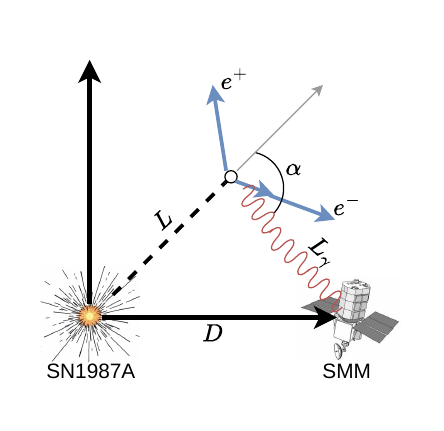}
    \captionsetup{justification=Justified}
    \caption{Illustration of the  geometry effect for the scalar emission from SN1987A and decay into lepton-antilepton pair, which produce secondary photons that would have led to an excess over the background at the SMM gamma-ray satellite.}
    \label{fig:geometry}
\end{figure}
The scalar travels a distance $L$ from the core of the supernova, then decays to either leptons or photons. The leptons can further give rise to secondary photons. All these photons then travel distance $L_\gamma$ to reach the experiment. The distance between the experiment and the supernova is $D$. This geometric description of the scalar event is shown in Fig.~\ref{fig:geometry}. For the direct photon decay, the three lengths $L$, $L_\gamma$ and $D$ form a triangle (see Fig. S6 of~\cite{Dev:2021kje}). Whereas for the secondary photon scenario, the above mentioned three distances do not make an exact triangle. However, the concerned distances are significantly huge and hence $L$, $L_\gamma$ and $D$ can be assumed to create an approximate triangle. Under the assumption of the triangle formation, $L_\gamma$ can be written in the following way:
\begin{equation}\label{Eq:Lgamma}
L_\gamma=-Lz_\alpha+\sqrt{D^2-L^2(1-z_\alpha^2)} \,.
\end{equation}
The Heaviside Theta function $\Theta(D/\sqrt{1-z_\alpha^2}-L)$ in Eq.~\eqref{Eq:mainprod} reflects the viability of Eq. \ref{Eq:Lgamma}.
\par The term, $(m_S^2)/(2\omega_S^2(1-\beta_S z_\alpha)^2)$, in Eq.~\eqref{Eq:mainprod} denotes the $\alpha$ angle distribution boosted to the Earth reference frame and the term, $1/(4\pi D(L_\gamma+Lz_\alpha))$, turn up in Eq.~\eqref{Eq:mainprod} from the requirement that only photons emitted with the appropriate angle, $\alpha$ can reach the detector. More details on these terms and the decay geometry can be found in~\cite{Dev:2021kje}.

\section{Secondary Photon Spectrum from a Single Scalar Decay}
\label{App:pythiamatch}
The secondary photon spectrum from the decay of scalar into lepton-antilepton pairs in its rest frame takes the following form~\cite{Fortin:2009rq} :
\begin{equation}
    \frac{dN_\gamma}{d\omega_\gamma}\Big|_\text{unboosted}\simeq \frac{\alpha}{\pi}\left( \frac{m_S^2+(m_S-2\omega_\gamma)^2}{m_S^2 \omega_\gamma}\text{ln}\left[\frac{m_S(m_S-2\omega_\gamma)}{m_l^2}\right]\right)
\label{Eq:unboostdNdEgamma}
\end{equation}
with $\alpha$ being the fine structure constant and $m_l$ being the mass of the lepton and antilepton that $S$ decays into. \par
\begin{figure*}[t!]
  \centering
  \includegraphics[width=0.48\textwidth]{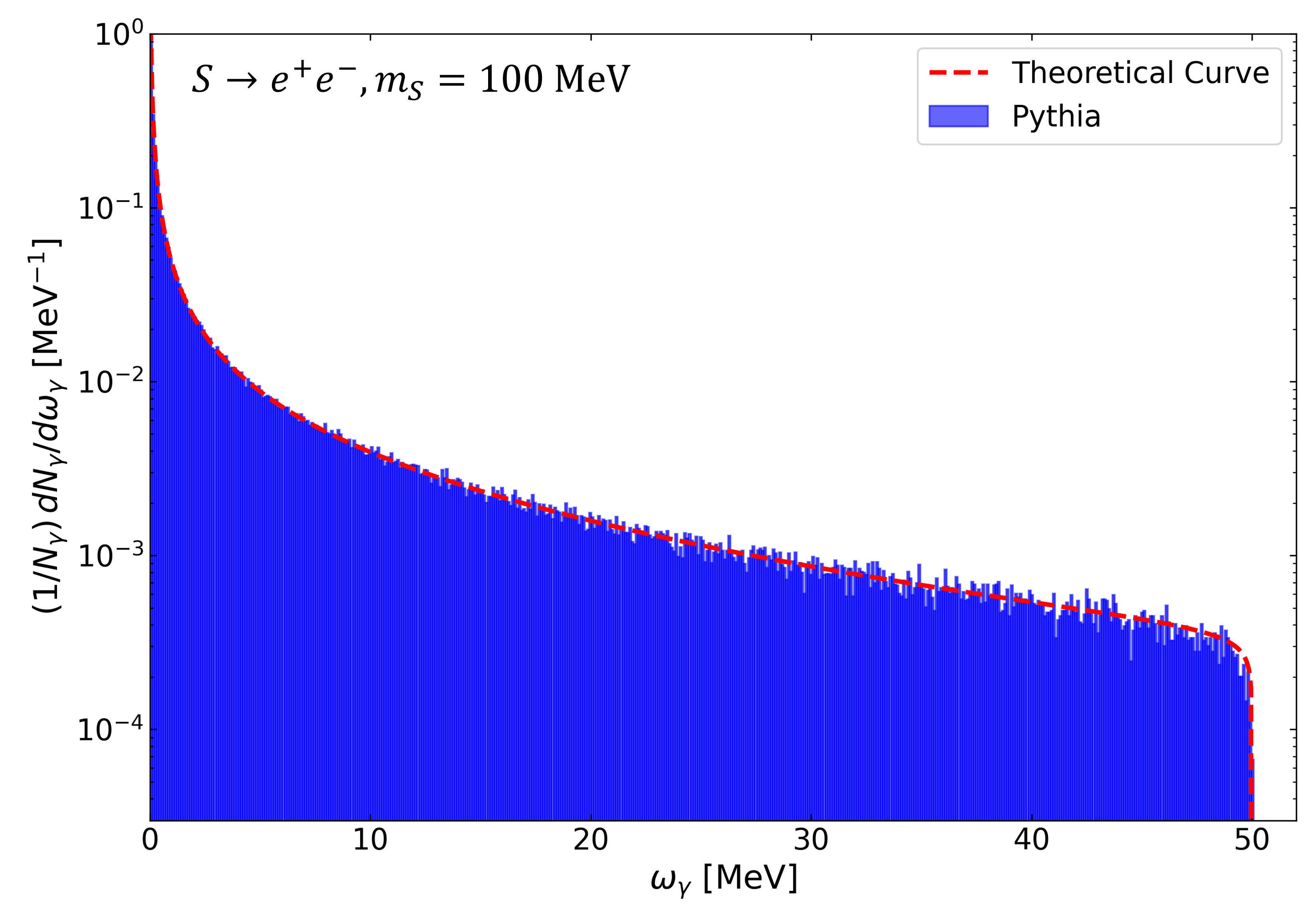}
  \hspace{0.02\textwidth}
  \includegraphics[width=0.48\textwidth]{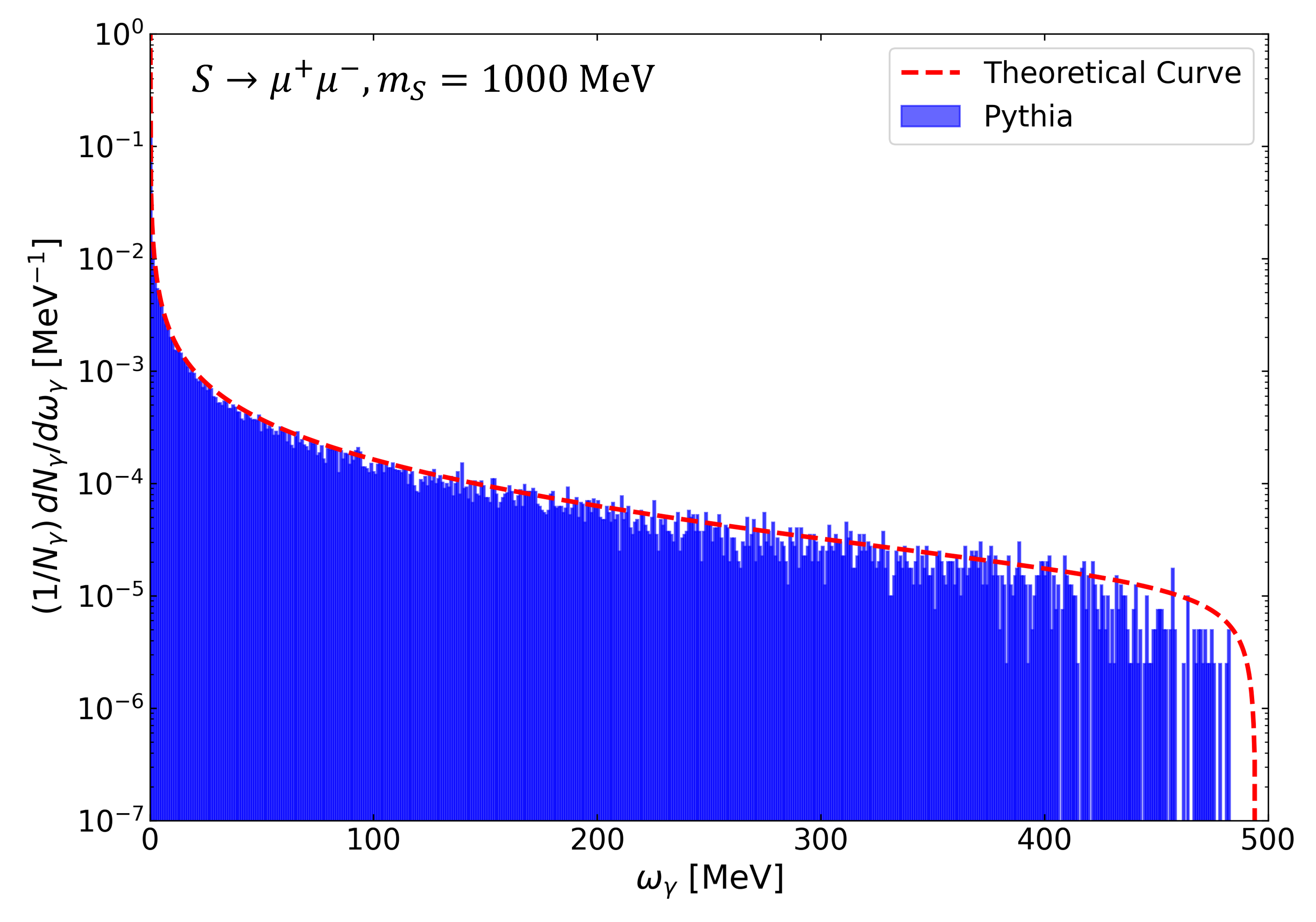}
  \captionsetup{justification=Justified}
  \caption{{\it Left:} the unboosted secondary photon spectrum from $S$ decaying into electronic channel. {\it Right:} the unboosted secondary photon spectrum from $S$ decaying into muonic channel. }
  \label{fig:pythiamatch}
\end{figure*}

 The unboosted secondary photon spectrum for both $S\rightarrow e^-e^+$ and $S\rightarrow \mu^-\mu^+$ is simulated using MadGraph~\cite{Alwall:2014hca} plus pythia~\cite{Sjostrand:2014zea} and that matches very well with the functional form in Eq.~\eqref{Eq:unboostdNdEgamma} (Fig.~\ref{fig:pythiamatch}).

\par For a scalar particle with energy $\omega_S$, the secondary photon spectrum from the scalar decay looks like:
\begin{equation}
    \frac{dN_\gamma}{d\omega_\gamma}\Big|_\text{boosted}=\int_{z^\text{CM}_\text{min}}^{1} \frac{dz^\text{CM}}{2}\frac{m_S}{\omega_S+z^\text{CM}\sqrt{\omega_S^2-m_S^2}}\frac{dN_\gamma}{d\omega_\gamma}\Big|_\text{unboosted}
\end{equation}
where, $z^\text{CM}_\text{min}$ is the maximum of $-1$ and $z^*$ with 
$$z^*=\frac{2\omega_\gamma}{\sqrt{\omega_S^2-m_S^2}(1-\frac{4m_l^2}{m_S^2})}-\frac{\omega_S}{\sqrt{\omega_S^2-m_S^2}}\,.$$ 
The expression of $z^*$ comes from the fact that in the rest frame of the scalar, there is a maximum energy that the secondary photon can have \cite{Fortin:2009rq}.

\twocolumngrid

\bibliographystyle{apsrev4-2}
\bibliography{ref}

\end{document}